\documentclass[aps,prb,unsortedaddress, %superscriptaddress,
               showpacs,footinbib,amsfonts,amssymb,amsmath,nobibnotes,
               twocolumn,
              ]{revtex4-1}
\usepackage[utf8]{inputenc}
\usepackage{amssymb}
\usepackage{amsmath}
\usepackage{amsthm}
\usepackage{mathrsfs}
\usepackage{graphicx}
\usepackage{color}
\usepackage{braket}
\usepackage{placeins}
\usepackage{mathptmx}
\usepackage{hyperref}
\hypersetup{colorlinks=true,linkcolor=blue,citecolor=blue,urlcolor=blue}
\usepackage{lipsum}
\usepackage{bm}
\usepackage{graphicx}

\graphicspath{{img/}{img/model/}{img/diagrams/}}

\renewcommand{\Re}{\text{Re}}
\renewcommand{\Im}{\text{Im}}

\newcommand{\mv}[1]{\mathbf{#1}}
\renewcommand{\vr}{\mv{r}}
\newcommand{\vR}{\mv{R}}
\newcommand{\vk}{\mv{k}}
\newcommand{\vq}{\mv{q}}
\newcommand{\vQ}{\mv{Q}}

\newcommand{\ieig}{i}
\newcommand{\jeig}{{i'}}
\newcommand{\eig}{\varepsilon}

\newcommand{\ipho}{\lambda}

\newcommand{\ispin}{\sigma}

\newcommand{\sym}{\mathfrak{S}}

\newcommand{\Ttau}{T_{\tau}}

\newcommand{\icell}{l}
\newcommand{\iat}{\kappa}
\newcommand{\icart}{j}
\newcommand{\jcell}{\icell'}
\newcommand{\jat}{\iat'}
\newcommand{\jcart}{\icart'}

\begin{abstract}

The effects of the electron-phonon interaction on optical excitations
  can be understood in terms of exciton-phonon coupling,
  and require a careful treatment
  in low-dimensional materials with strongly bound excitons
  or strong electron-hole interaction in general.
Through phonon absorption and emission processes,
  the optically accessible excitons are scattered
  into otherwise optically dark
  finite-momentum exciton states.
We derive a practical expression for the
  phonon-induced term of the exciton self-energy
  (denoted as the exciton-phonon self-energy)
  that gives the temperature dependence
  of the optical transition energies and their lifetime broadening
  resulting from the exciton's interaction with the phonons.
We illustrate this theory on a two-dimensional model,
  and show that our expression for the exciton-phonon self-energy
  differs qualitatively
  from previous expressions found in the literature
  that neglect the exciton binding or electron-hole correlations.

\end{abstract}

\begin{document}

\title{Theory of Exciton-Phonon Coupling}
\author{Gabriel Antonius}
\email{gabriel.antonius@uqtr.ca}
\affiliation{D\'epartement de Chimie, Biochimie et Physique, Institut de recherche sur l'hydrog\`ene, Universit\'e du Qu\'ebec \`a Trois-Rivi\`eres, Trois-Rivi\`eres, C.P. 500, Trois-Rivières, Canada G9A 5H7}
\author{Steven G. Louie}
\affiliation{Department of Physics, University of California at Berkeley, California 94720, USA, and Materials Sciences Division, Lawrence Berkeley National Laboratory, Berkeley, California 94720, USA}
\maketitle

A wealth of two-dimensional and nanocrystalline materials
  with interesting optical properties
  have been studied in recent years,
  including
  transition-metal dichalcogenides, %monochalcogenides,
  layered heterostructures
  and halide perovskites~%
  \cite{%
  bhimanapati_recent_2015,%   All 2D materials
  mueller_exciton_2018,%   TMD
  schaibley_valleytronics_2016,%   valleytronics, heterostructures
  novoselov_2d_2016,%   heterostructures
  kovalenko_properties_2017,%   perovskite nanocrystals
  shi_two-dimensional_2018,%   2d perovskite and other 2d materials
  chen_two-dimensional_2017%   2d perovskite etc.
  }%
  .
In these systems, the optical excitations
  lead to the formation
  of strongly bound electron-hole
  pair states known as excitons.
Many of their useful
  opto-electronic properties (e.g., photocurrent generation, single-photon emission, etc.)
  depend on the scattering dynamics and diffusion of the excitons
  \cite{%
  shi_exciton_2013,% Exciton Dynamics in suspended MoS2
  yuan_exciton_2017,%  Exciton Dynamics, Transport, and Annihilation
  kozawa_photocarrier_2014,%  Photocarrier relaxation pathway (xct-xct)
  ruppert_role_2017,%  Role of phonons in exciton diffusion
  raja_enhancement_2018,%  Exciton-phonon scattering in WS2
  li_phonon-suppressed_2019,%  Phonon-suppressed Auger Scattering
  chi_ultrafast_2018%  Energy dissipation due to phonons in MoS2
  }%
  .
This dynamics is governed by several processes:
  the interaction of excitons with defects,
  the exciton-exciton interaction,
  and the exciton-phonon interaction.
In particular, exciton-phonon coupling effects
  can be identified by their distinctive temperature dependence,
  whether in the
  exciton mean free path,
  lifetimes or emission spectra
  \cite{%
  selig_excitonic_2016,%  Temperature dependence of the linewidth in WS2, MoSe2
  gong_electronphonon_2018,%  Perovskite blue emitters
  lv_exciton-acoustic_2021%  Satellite peaks in PL spectra in perovskites
  }%
  .

The exciton-phonon coupling mechanism originates from
  a combined action of
  the electron-hole and the electron-phonon interactions,
  both of which can be described from first principles.
On the one hand,
 the electron-hole interaction underlies the formation of excitons,
 and can be addressed %computed from first principles
 with the Bethe-Salpeter equation (BSE)
 within the \textit{ab initio} $GW$-BSE method~%
 \cite{%
  strinati_effects_1984,%
  strinati_application_2008,%
  Albrecht1998,%
  benedict_theory_1998,%
  rohlfing_excitonic_1998,%
  rohlfing_electron-hole_1998,%
  Rohlfing2000,%
  Onida2002%
  }.
This method solves the interacting
  two-particle problem for an electron and a hole,
  and yields the exciton energies and wavefunctions,
  which allow to predict the optical absorption spectra of materials
The electron-phonon interaction, on the other hand,
  has largely been studied within
  density functional perturbation theory (DFPT)~%
  \cite{Gonze1997,Gonze1997a,Baroni2001},
  which provides an \textit{ab initio} description of the phonon energy spectrum
  and coupling potential.
This framework has been used to study
  the effect of phonons
  on the band structure and carrier mobility
  as a function of temperature
  \cite{%
  giustino_electron-phonon_2017,%
  miglio_predominance_2020,%
  ponce_first-principles_2020%
  }%
  .
Going beyond DFT, electron-electron correlation effects
  to the electron self-energy may further be included
  at the $GW$ level from \textit{ab initio} \cite{Hybertsen1986}
  using the $GW$PT method \cite{li_electron-phonon_2019}
  in computing the electron-phonon interaction.

Describing the dynamics of photoexcited states from first principles
  is a challenging task.
Simulations of hot electrons
  were achieved by retaining the electron-phonon interaction only,
  with the rationale that electrons far from the band edges
  would scatter freely without forming bound excitons
  \cite{%
  bernardi_textitab_2014,%
  molina-sanchez_ab_2017,%
  caruso_nonequilibrium_2021%
  }%
  .
It is necessary, however, to include the electron-hole interaction
  to predict the lifetime of absorption and emission states
  when they originate from bound excitons.

An early attempt to compute the temperature-dependent broadening
  and renormalization of exciton states
  was based on a one-particle picture of the electron-phonon coupling~%
  \cite{Marini2008}.
This scheme has been used to compute the
  absorption spectrum of h-BN and MoS$_2$ at finite temperature
  \cite{Marini2008,qiu_optical_2013, molina-sanchez_temperature-dependent_2016}.
In this approach, the electron-phonon renormalization and broadening
  of the band structure is computed before solving the BSE. 
This method does not, however, describe correctly the process where excitons
  scatter into finite-momentum bound states,
  which is necessary to enforce energy conservation.
Alternatively,
  the supercell BSE technique~%developed by M. Zacharias et al~%
  \cite{%
  zacharias_stochastic_2015,%
  zacharias_one-shot_2016,%
  zacharias_theory_2020%
  }
  does account for the phonon-mediated interaction
  between optical excitons and
  a limited number of 
  finite-momentum excitons
  commensurate to the supercell size.
It does so only within a static approximations,
  which is valid for non-polar materials.
This approach predicts the exciton energy renormalization
  as a function of temperature,
  but makes no prediction on the scattering lifetime of the excitons.
Recent methods formally 
  achieved a proper description of exciton dynamics
  with exciton-phonon scattering amplitude
  deduced from Fermi's golden rule
  using exciton-phonon coupling matrix elements
  \cite{%
  brem_exciton_2018,%
  brem_intrinsic_2019,%
  christiansen_theory_2019,%
  chen_exciton-phonon_2020%
  }%
  .
This approach enforces energy conservation,
  and is consistent with the theory presented in this paper,
  as well as other methods derived from many-body perturbation theory~%
  \cite{cudazzo_first-principles_2020}.
Another effect of phonons on the exciton binding energies
  comes from the lattice screening,
  and this has been recently computed from first principles~%
  \cite{filip_phonon_2021,adamska_bethe-salpeter_2021}

In this work, we develop a general theory of the exciton-phonon coupling
  that is amenable to first-principles calculation.
The central object is the exciton-phonon self-energy
 (i.e., the contribution to the exciton self-energy
  due to exciton-phonon interaction;
  there are of course contributions due to other excitations in a system),
  which yields the energy renormalization of the exciton
  states, as well as their scattering lifetime due to phonons.
We apply this theory
  to a tight-binding model
  in two dimensions
  and discuss how it differs from other methods.
This paper is organized as follows.
Section~\ref{sec:ehepint}
  reviews the theory of electron-hole and electron-phonon interactions.
Section~\ref{sec:TdepAbsorption}
  presents the extension of the one-particle theory
  to the exciton-phonon coupling,
  and the main equations for the
  phonon-induced temperature-dependent 
  exciton lifetimes and energies.
In Sec.~\ref{sec:Application},
  we apply this scheme to a 2D tight-binding model,
  and discuss the consequences of electron-hole interactions
  on the scattering dynamics of two-dimensional systems.
The main findings are summarized in Sec.~\ref{sec:Conclusion}.
Several mathematical details of the derivation
  can be found in supplemental materials~\cite{SM}.

\section{Electron-hole and electron-phonon interactions}
\label{sec:ehepint}

As a starting point for the treatment of electron-hole
  and electron-phonon interactions,
  we consider a mean-field fixed-ions Hamiltonian for the electrons
  $H_0 = \hat T_k + V^{SCF}(\vr)$,
  where $\hat T_k$ is the kinetic energy operator,
  and $V^{SCF}(\vr)$ is the self-consistent field potential.
Within the density functional theory (DFT) framework,
  $V^{SCF}$ includes the potential of the ions, the Hartree potential
  and the exchange-correlation potential. 
Solving the one-particle Hamiltonian yields the set of unperturbed
  wavefunctions $\phi_{\ieig}(\vr)$
  and energies $\eig_\ieig$, where the label $\ieig$
  comprises a band index ($n_\ieig$), a wavevector ($\vk_\ieig$),
  and eventually a spin index ($\sigma_\ieig$).

These quantities are used to construct the time-ordered Green's function,
  defined in the one-particle basis and in time as
  $G^0_{\ieig\jeig}(t)\!=\!-i \braket{\hat T_t c_\ieig (t) c_\jeig^\dag(0)}_0$,
  where $c_\ieig^\dag$ and $c_\ieig$ are the electron
  creation and annihilation operators,
  $\hat T_t$ is Wick's time-ordering operator,
  and "$0$" indicates here that the expectation
  value is taken over a ground state
  that is not perturbed by the phonons.
The creation and annihilation operators follow the commutation relations
\begin{equation} \label{eq:commut}
  \big\{ c_{\ieig}, c_{\jeig}^\dag \big\}
  = c_{\ieig} c_{\jeig}^\dag + c_{\jeig}^\dag c_{\ieig}
  = \delta_{\ieig\jeig}
\end{equation}
and $\{ c_{\ieig}, c_{\jeig}\} = \{ c_{\ieig}^\dag, c_{\jeig}^\dag\}=0$.
In terms of frequencies, the one-particle Green's function writes as
\begin{equation}
  G^0_{ii'}(\omega) = \frac{1}{\omega - \eig_i \pm i \eta }
                      \delta_{ii'}
\end{equation}
where $\pm\eta$ is an real infinitesimal number
  with the same sign as $\eig_i$,
  the eigenvalue measured with respect
  to the chemical potential.

The electronic energies and wavefunctions
  that define the starting point need not be obtained from DFT;
  they may also be obtained from a model hamiltonian
  or from a many-body scheme.
It is required however to have an effective Hamiltonian
  that depends implicitly on the atomic coordinates
  and that can be differentiated with respect to these coordinates
  to obtain the dynamical matrix and the electron-phonon coupling elements.
Such a Hamiltonian, for example, may be constructed from the self-energy
  computed in the $GW$ formalism \cite{vanSchilfgaarde2006,Antonius2014,li_electron-phonon_2019}.

\subsection{Electron-hole interaction}
A class of
  neutral excitations
  of an insulating system
  is composed of an electron being
  promoted into the conduction bands and leaving a hole in the valence bands.
If the Coulomb attraction between the electron and the hole
  is sufficiently strong, they may form a bound exciton,
  that is, a bound state whose excitation
  energy is smaller than
  the fundamental band gap.
The procedure to compute the exciton spectrum from the
  BSE is described in Ref.~\onlinecite{Rohlfing2000}.

The starting point to describe excitons is the set of 
  independent (or non-interacting) electron-hole pairs,
  typically with quasiparticle energies from a $GW$ calculation
  in the \textit{ab initio} $GW$-BSE approach.
The corresponding propagator for these
  fictitious excitations is
\begin{align}
  L^0_{vcv'c'}(\omega) = \frac{1}{\omega - (\eig_c - \eig_v) + i\eta}
                         \delta_{cc'} \delta_{vv'}
\end{align}
where $v$ and $c$ refer to
  the labeling of 
  occupied (valence band)
  and unoccupied (conduction band) states
  which involve both band and $\vk$-point indices.
We call $L^0$ the bare electron-hole propagator.
It is computed with the fixed-ions Hamiltonian,
  but the superscript "$0$" refers to the fact that the
  electron and hole propagate freely without interacting with one another.

\begin{figure}[t]
  \includegraphics[width=0.8\linewidth]{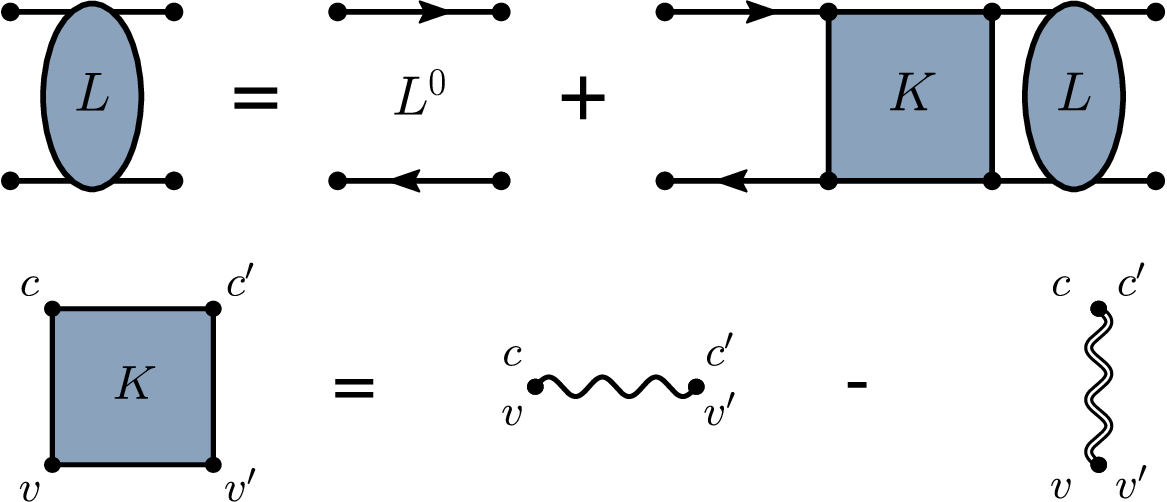}
  \caption{\label{fig:BSE}
    Diagramatic representation of the Bethe-Salpeter equation.
    The BSE Kernel is expressed as the sum of the bare exchange Coulomb repulsion (single line)
    and the screened Coulomb attraction (double line) between the electron and the hole.
    }
\end{figure}

The BSE relates the bare exciton propagator $L$
  to the bare electron-hole propagator $L^0$
  through the kernel $K$,
  as depicted in Fig.~\ref{fig:BSE}.
The BSE kernel $K$ is composed of an attractive screened Coulomb
  interaction between the electron and the hole,
  and a repulsive Coulomb exchange term.
In practice, 
  the BSE is solved by diagonalizing an effective
  two-particle Hamiltonian,
\begin{equation}
  H^{2p}_{vc,v'c'} = (\eig_c - \eig_v) \delta_{cc'} \delta_{vv'} + K_{vc,v'c'}
\end{equation}
  where $K_{vc,v'c'}$ is the static version of the BSE kernel,
  and the first term is the sum of the quasiparticle energies
  of the electron and hole. 
This Hamiltonian yields the exciton energies $\Omega_S$
  and electron-hole coefficients $A_{vc}^S$
  for each exciton $S$.
The resonant part of the bare exciton propagator
  in the quasiparticle basis 
  can thus be written as
\begin{align}
  L_{vcv'c'}(\omega) = \sum_S \frac{A_{vc}^S A_{v'c'}^{S*}}
                                       {\omega - \Omega_S + i\eta}
.
\end{align}
Like the one-particle state indices,
  the label $S$ comprises
  a set of discrete quantum numbers as well as the
  center-of-mass
  momentum of the exciton ($\mv q_S$)
  which is the wavevector difference ($\mv k_c - \mv k_v$)
  between the unoccupied and occupied 
  single-particle orbitals
  that form the exciton.
  (The periodicity of the crystal dictates that all free electron-hole pairs
  forming the exciton have the same wavevector difference
  modulo a reciprocal lattice vector.) 
Because of the small momentum carried by photons,
  only excitons with $\mv q_S\approx0$ are optically accessible.
We will see that finite-momentum excitons ($\mv q_S \neq 0$) are important
  to describe scattering events by phonons.

Since the $A_{vc}^S$ are eigenfunctions of the two-particle Hamiltonian,
  they form a complete orthonormal basis for the space spanned by
  the set of valence and conduction bands used to construct $H^{2p}$.
We can use these coefficients to transfer between the $vc$ basis
  and the $S$ basis.
For example, the bare exciton propagator writes
\begin{align} \label{eq:basistransform}
 L_{SS'}(\omega) =
  \sum_{vc,v'c'} A_{vc}^{S*} A_{v'c'}^{S'} L_{vcv'c'}(\omega)
  = \frac{1}{\omega - \Omega_S + i\eta}\delta_{SS'}
.
\end{align}

The absorption spectrum at zero temperature
  without electron-phonon interaction
  can be constructed from the exciton energies and electron-hole coefficients.
It takes the form
\begin{align} \label{absp}
  \epsilon''(\omega)
    = \frac{4\pi^2 e^2}{\omega^2}
    \sum_{S}
    \big\vert
      \sum_{vc} A_{vc}^{S} \bra{v} \mv e \cdot \mv v \ket{c}
    \big\vert^2
    \delta(\omega-\Omega_S)
\end{align}
where $\mv v$ is the velocity operator, $\mv e$ is the photon's
  polarization vector,
  and the summation over exciton states is restricted to zero-momentum
  excitons.
The delta function in Eq.~\eqref{absp} is typically
  represented as a Lorentzian function
  with a certain broadening. % parameter $\eta$.
In most past calculations,
  this broadening was chosen empirically to reproduce
  the available experimental data.
The absorption line broadening is in fact related to the lifetime
  of the excitons and, as we show, can be computed from first principles.

\subsection{Lattice dynamics and electron-phonon interaction}

The lattice dynamics can be obtained from a self-consistent calculation
  of the dynamical matrix of the crystal ($\Phi$),
  as detailed in Ref.~\onlinecite{Gonze1997a}.
In real space,
  the dynamical matrix (or force matrix)
  corresponds to the second-order derivative of
  the total energy with respect to the displacement of two atoms:
  $\Phi^{\iat\jat}_{\icart\jcart} (\mv R_{\icell} - \mv R_{\jcell})
  = \nabla_{\icell\iat\icart} \nabla_{\jcell\jat \jcart} E$,
  where $\icell$ labels a unit cell of the crystal
  with lattice vector $\mv R_{\icell}$,
  $\iat$ labels an atom
  in
  the unit cell,
  and $\icart$ labels a Cartesian direction.
The phonon frequencies $\omega_{\ipho}$
  and polarization vectors $U^{\ipho}_{\iat\icart}$,
  are obtained by diagonalizing the Fourier transform of $\Phi$ as
\begin{align}
  M_{\iat} \omega_{\ipho}^2 U^{\ipho}_{\iat\icart}
    = \sum_{\jat, \jcart} \Phi^{\iat\jat}_{\icart\jcart} (\mv{q})
      U^{\ipho}_{\jat\jcart}
\end{align}
  where $M_{\iat}$ is the mass of an atom.
The label $\ipho$ for a phonon mode comprises
  a branch index and a wavevector ($\vq$ or $\vq_\ipho$).
The DFPT method allows for the self-consistent calculation of
  the first-order derivative of the local potential
  with respect to atomic displacements,
  $\nabla_{\icell\iat\icart} V^{SCF}(\vr)$.
Thanks to the Hellman-Feynman theorem and the $2n+1$ theorem,
  only the first-order derivatives
  of the potential and density need to be computed self-consistently
  in order to construct the dynamical matrix \cite{gonze_density-functional_1989}.

The phonon propagator is defined in time as %in the $\ipho$ basis as
  $D_{\vq\ipho}(t)\!=\!-i\braket{\hat T_t A_{\vq \ipho} (t) A_{-\vq \ipho} (0)}_0$,
  where $A_{\vq\ipho}\!=\!a^{\dagger}_{\vq\ipho}\!+\!a_{\vq\ipho}$ is the sum
  of a phonon creation and annihilation operator.
In frequency space, the phonon propagator writes as
\begin{equation}
  D_{\ipho}(\omega) = \frac{1}{\omega - \omega_{\ipho} + i\eta}
                  - \frac{1}{\omega + \omega_{\ipho} + i\eta}
.
\end{equation}

The electron-phonon interaction stems from
  the perturbation to the fixed-atoms Hamiltonian
  created by the phonons.
A thorough discussion of the electron-phonon interaction
  in the context of DFT is presented in Ref.~\onlinecite{marini_manybody_2015,giustino_electron-phonon_2017}.
Expanding the Hamiltonian up to second order in the atomic displacements,
  the electron-phonon coupling potential writes as
\begin{align} \label{Vep}
  V_{ep} = V^{(1)}_{ep} + V^{(2)}_{ep}
         = \sum_{ii' \ipho} g_{ii' \ipho}
              A_{\ipho} c^{\dagger}_{i} c_{i'}
         + \sum_{ii' \ipho \ipho'}
               g^{(2)}_{ii' \ipho \ipho'}
              A_{\ipho} A_{\ipho'} c^{\dagger}_{i} c_{i'}
\end{align}
with the first-order electron-phonon coupling matrix elements
\begin{align} \label{gkk}
  g_{ii' \ipho}(\vk,\vq) = \sqrt{\frac{\hbar}{2M\omega_{\vq\ipho}}}
    \Bra{\phi_{i\vk+\vq}} \nabla_{\vq\ipho} V^{SCF}(\vr) \Ket{\phi_{\vk i'}}
\end{align}
and the second-order matrix elements 
\begin{align} \label{gDW}
   g^{(2)}_{ii' \ipho \ipho'}(\vk,\vq,\vq') = &
    \frac{\hbar}{2M\sqrt{\omega_{\vq\ipho} \omega_{\vq'\ipho'}}}\nonumber\\
    & \times \Bra{\phi_{i\vk+\vq+\vq'}} \tfrac{1}{2} \nabla_{\vq\ipho}
      \nabla_{\vq'\ipho'} V^{SCF}(\vr) \Ket{\phi_{i'\vk}}
.
\end{align}
The derivative of the potential with respect to a phonon mode
  is defined as
\begin{equation}
  \nabla_{\vq\ipho} V^{SCF}(\vr)
    = \sum_{\icell\iat\icart} U^{\ipho}_{\iat\icart}(\vq)
        e^{i \vq \cdot \mv R_{\icell}}
        \nabla_{\icell\iat\icart} V^{SCF}(\vr)
.
\end{equation}

The wavevectors ($\vk, \vq$) are written explicitly in  Eq. \eqref{gkk} and \eqref{gDW},
  but will be omitted
  in the remainder of the paper 
  to lighten the notation.
Also, to make the units explicit, % in Eq. \eqref{gkk} and \eqref{gDW},
  we %wrote the values of
  use 
  $\hbar=1$
  and the mass $M$,
  which normalizes the phonon eigenvectors according to
  $\sum_{\iat} M_{\iat} \vert U^{\ipho}_{\iat}\vert^2 = M$.
It
  is useful to assign $M$ the value
  of the average mass of the atoms of the unit cell,
  so that the phonon eigenvectors are on the order of unity,
  and the factor $1/M$ serves as an expansion parameter.
Note that each summation over phonon modes in Eq. \eqref{Vep}
  is implicitly normalized by $\sqrt{N_{\vq}}$,
  where $N_{\vq}$ is the number of wavevectors
  used to sample the Brillouin zone.

The quantity $g_{ii' \ipho}$ is the electron-phonon coupling
  matrix element between the one-particle states $i$ and $i'$
  via the phonon mode $\ipho$.
Because of crystal momentum selection rule,
  for $g_{ii' \ipho}$ to be non-zero,
  we must have
  $\mv k_{i} = \mv k_{i'} + \mv q_{\ipho} + \mv G$
  where $\mv G$ is a reciprocal lattice vector
  that is non-zero for an Umklapp process.
For $g^{(2)}_{ii' \ipho \ipho'}$,
  at the lowest order of perturbation theory,
  we must have $\mv k_{i} = \mv k_{i'}$
  and $\vq_{\ipho'}=-\vq_{\ipho}$ for the wavevectors,
  and $\lambda'=\lambda$ for the branch indices.
In practice, the second-order electron-phonon coupling matrix elements
  $g^{(2)}_{ii' \ipho \ipho}$ are never computed explicitly.
Their contribution to the self-energy is approximated
  in terms of the first-order electron-phonon coupling matrix elements
  by making use of the accoustic sum rule 
  and the rigid-ion approximation~\cite{Allen1976, Ponce2014}.

\subsection{One-particle propagators}

We employ the Matsubara formalism
  to treat the propagators and self-energies at finite temperature.
The propagators are defined on the imaginary time axis $\tau=it$
  in the interval $[-\beta,\beta]$ with $\beta=1/k_B T$,
  and are made periodic outside of this range.
The one-particle propagator $G$
  is composed of odd (fermionic) Matsubara frequencies,
  while the even frequencies compose
  all bosonic propagators:
  electron-hole ($L^0$), exciton ($L$), and phonon ($D$).

\begin{figure}[t]
  \includegraphics[width=0.8\linewidth]{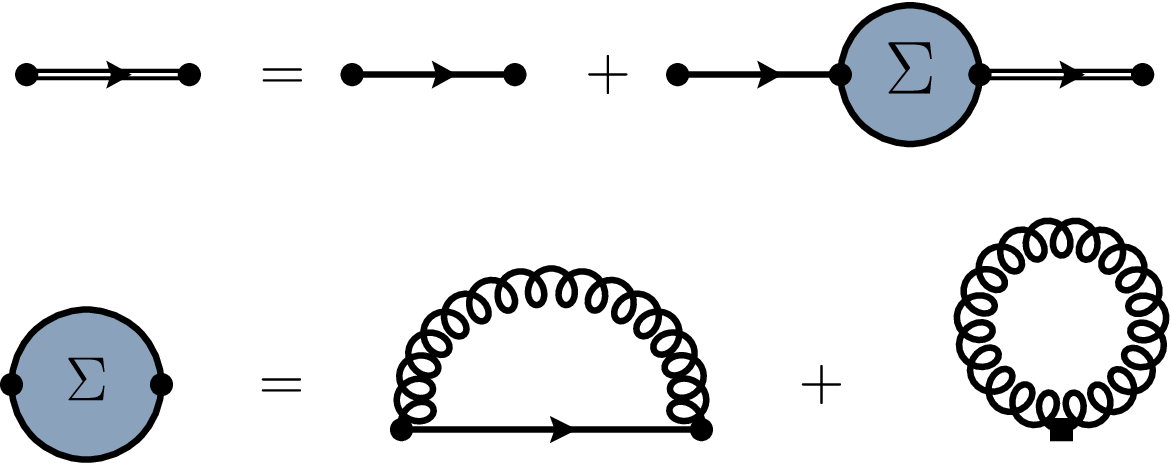}\\
  \caption{\label{fig:Dyson_G}
    Dyson equation for the one-particle propagator
    and the electron-phonon self-energy,
    expressed as the sum of the Fan-Migdal term
    and the Debye-Waller term.
    }
\end{figure}

The interacting Green's function $G$ can be expanded
  in powers of the perturbation as
\begin{align} \label{Gtauexpanded}
  G_{ii'}(\tau) & = - \sum_{n=0}^{\infty} (-1)^n
    \int_{0}^{\beta}d\tau_1 \dots \int_{0}^{\beta}d\tau_n\nonumber\\
  & \qquad \times \braket{\Ttau c_{i}(\tau) c^{\dagger}_{i'}(0)
      V_{ep}(\tau_1)\dots V_{ep}(\tau_n)}_0
\end{align}
and we adopt the convention that 
  in each expectation value ${\langle \dots \rangle}$,
  only distinct and connected diagrams should be retained~%
  \cite{Mahan2000}.
The time-dependent operators are expressed in the interaction
  picture, that is
\begin{equation}
  c_{i}(\tau) = e^{\tau \hat H_0} c_{i}(0) e^{-\tau \hat H_0}
\end{equation}
with the mean field Hamiltonian ${\hat H_0 = \sum_i \eig_i c^{\dagger}_i c_i}$.
Equation \eqref{Gtauexpanded} can be cast into a Dyson equation for $G$,
  depicted in Fig.~\ref{fig:Dyson_G}, as
\begin{align} \label{eq:Dyson_G}
  G_{ii'}(\omega, T) & = G^0_{ii'}(\omega)
   + \sum_{i_1 i_2} G^0_{ii_1}(\omega)
                    \Sigma_{i_1i_2}(\omega, T)
                    G_{i_2i'}(\omega, T)
\end{align}
  which defines the one-particle electron-phonon self-energy $\Sigma$.

The perturbative expansion of $G$ with Eq. \eqref{Gtauexpanded}
  yields powers of $V_{ep}^{(1)}$ and $V_{ep}^{(2)}$,
  where $V_{ep}^{(1)}$ appears an even number of times in each term.
The Migdal theorem %~\cite{Migdal1958}
  states that this expansion
  can be truncated to the lowest power of $1/M$.
Since $V_{ep}^{(1)}$ is proportional to $1/\sqrt{M}$ and
  $V_{ep}^{(2)}$ is proportional to $1/M$,
  two terms remain in the self-energy after truncation:
  the Fan-Migdal (FM) term and the Debye-Waller (DW) term, written as
\begin{equation}
  \Sigma_{ii'}(\omega, T) = \Sigma^{\text{FM}}_{ii'}(\omega,T) + \Sigma^{\text{DW}}_{ii'}(T)
.
\end{equation}
The FM term is dynamic (i.e. complex and frequency-dependent)
  and stems from the first-order
  electron-phonon coupling potential. % appearing twice. %, giving
Its analytic expression is
\begin{align} \label{Fanonep}
  \Sigma^{\text{FM}}_{ii'}(i\omega_n) = &
    \sum_{i'' \ipho}
    g_{ii'' \ipho} g^{*}_{i' i'' \ipho}\nonumber\\
    & \times
    \Big[ \frac{N_B(\omega_\ipho) + f(\eig_{i''})}
               {i\omega_n - \eig_{i''} + \omega_\ipho}
        + \frac{N_B(\omega_\ipho) + 1 - f(\eig_{i''})}
               {i\omega_n - \eig_{i''} - \omega_\ipho}
    \Big]
\end{align}
where $N_B(\omega)$ is the Bose-Einstein distribution,
  and $f(\omega)$ is the Fermi-Dirac distribution,
  both of which depend implicitly on temperature.
The self-energy is evaluated on the real frequency axis
  with the analytic continuation $i\omega_n\rightarrow \omega + i\eta$
  where $\eta$ is an infinitesimal real number that is taken
  positive for the retarded Green's function (%
  for unoccupied states
  or electrons%
  )
  and negative for the advanced Green's function (%
  for occupied states
  or holes%
  ).

The Debye-Waller term is static (real and frequency-independent),
  and it stems from the second-order electron-phonon coupling potential.
It writes
\begin{align} \label{DWonep}
  \Sigma^{\text{DW}}_{ii'}
    = \sum_{\ipho} g^{(2)}_{ii' \ipho \ipho} \big[2N_B(\omega_\ipho) + 1\big]
.
\end{align}

The self-energy allows for the mixing of electronic states
  through the off-diagonal components of the self-energy ($i\neq i'$).
This mixing is only possible among states with 
  the same crystal momentum ($\vk_i=\vk_{i'}$).
If the bands are well-separated in energy,
  one may use only the diagonal elements of the self-energy to obtain
  the renormalization of the bands as
\begin{equation} \label{TempRenormalization}
  \eig_i + \Delta \eig_i(T) = \eig_i + \Re \Sigma_{ii}(\eig_i, T)
\end{equation}
and in general, $\Delta \eig_i(0) \neq 0$.
Correspondingly, the inverse lifetime of the
  electronic state is
\begin{equation} \label{Lifetime}
  \tau^{-1}_i(T) = 2 \vert \Im \Sigma_{ii}(\eig_i, T) \vert
.
\end{equation}
In Eq.~\eqref{TempRenormalization} and~\eqref{Lifetime},
  the temperature renormalization and lifetime is computed
  in the on-the-mass-shell limit, that is, by evaluating the self-energy at
  the bare energy, rather than the renormalized energy.
This is the preferred approach for phonon perturbations, %~\cite{Mahan2000},
  as it gives results in good agreement with a self-consistent
  calculation of the self-energy~\cite{brown-altvater_band_2020}.

\section{Exciton-Phonon self-energy: Self-energy of excitons from coupling to phonons}
\label{sec:TdepAbsorption}

We seek similar expressions for the
  exciton energy ($\Omega_S$) and lifetime ($\tau_S$)
  due to coupling to phonons 
  that would allow one
  to compute absorption spectra
  at zero and finite temperature through Eq.~\eqref{absp}.
The phonon-induced corrections will be given by the diagonal components
  of the exciton-phonon self-energy ($\Xi_{SS}$)
  to be discussed below, namely,
\begin{align} \label{OmegaRen}
  \Omega_S + \Delta \Omega_S(T) 
    & = \Omega_S + \Re \Xi_{SS}(\Omega_S, T)
\end{align}
for the exciton energies,
and
\begin{equation} \label{OmegaBrd}
  \tau_S^{-1}(T)= 2 \vert \Im \Xi_{SS} (\Omega_S, T) \vert
\end{equation}
  for the inverse lifetime of the excitons,
  that is, the absorption line broadening.

The expression for the exciton-phonon self-energy 
  is obtained by considering the
  interacting exciton propagator ($\Lambda$)
  defined as
\begin{align} \label{Lambdaexpandone}
  \Lambda_{vc,v'c'}(\tau)
  & =
  -\sum_{n=0}^{\infty} (-1)^n
    \int_{0}^{\beta}d\tau_1 \dots \int_{0}^{\beta}d\tau_n\nonumber\\
    \quad \times
  &
    \braket{\Ttau c_{c}(\tau^+) c^{\dagger}_{v}(\tau)
                  c_{v'}(0^+) c^{\dagger}_{c'}(0)
      \hat H_1(\tau_1)\dots \hat H_1(\tau_n)}_0
\end{align}
where $\tau^+$ is a time infinitesimally larger than $\tau$,
  and the interaction term $\hat H_1(\tau)$ is the sum of the
  static electron-hole interaction kernel $\hat K$
  and the electron-phonon coupling potential $\hat V_{ep}$.
Equation~\eqref{Lambdaexpandone} reduces to the BSE if $\hat V_{ep} = 0$.

The exciton-phonon self-energy connects the
  interacting exciton propagator $\Lambda$
  to the bare exciton propagator $L$
  (without electron-phonon interaction)
  in a Dyson-like equation:
\begin{equation} \label{LambdaDyson}
  \Lambda_{SS'}(\omega,T) = L_{SS'}(\omega)
    + \sum_{S_1S_2} L_{SS_1}(\omega)
        \Xi_{S_1S_2}(\omega,T) \Lambda_{S_2S'}(\omega,T)
.
\end{equation}
In order to express the exciton-phonon self-energy in an analytic
  form, it is useful to first consider
  the case without the electron-hole interaction.

\subsection{Independent electron-hole-pair-phonon self-energy}

\begin{figure}[t]
  \includegraphics[width=0.8\linewidth]{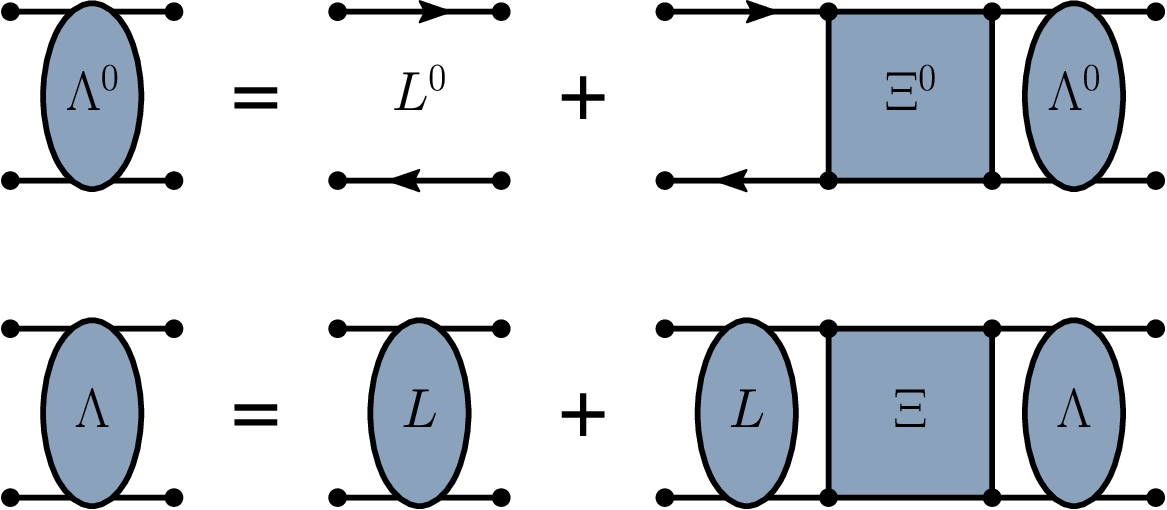}\\
  \caption{\label{fig:Dyson_Lambda}
    The Dyson equation for the
    IEHPP
    $\Lambda^0$
    involving the
    IEHPP
    self-energy $\Xi^0$
    (without electron-hole interactions),
    and the Dyson equation for the interacting
      exciton propagator $\Lambda$
      involving the exciton-phonon self-energy $\Xi$.
    }
\end{figure}

We define $\Lambda^0$
  as the electron-hole propagator with the electron-phonon interaction,
  but without the electron-hole interaction,
  calling it the independent electron-hole-pair-phonon (IEHPP) propagator,
  given by
\begin{align} \label{Lambdazexpandone}
  \Lambda^0_{vc,v'c'}(\tau)
  & =
  -\sum_{n=0}^{\infty} (-1)^n
    \int_{0}^{\beta}d\tau_1 \dots \int_{0}^{\beta}d\tau_n\nonumber\\
    \quad \times
  & \braket{\Ttau c_{c}(\tau^+) c^{\dagger}_{v}(\tau)
                  c_{v'}(0^+) c^{\dagger}_{c'}(0)
      \hat V_{ep}(\tau_1)\dots \hat V_{ep}(\tau_n)}_0
.
\end{align}
The corresponding
  IEHPP
  self-energy $\Xi^0$ is defined through
  the Dyson equation
\begin{align} \label{LambdazDyzon}
  \Lambda^0_{vc,v'c'}(\omega,T) & = L^0_{vc,v'c'}(\omega)\nonumber\\
    +
  & \sum_{v_1c_1v_2c_2} L^0_{vc,v_1c_1}(\omega)
                        \Xi^0_{v_1c_1v_2c_2}(\omega,T)
                        \Lambda^0_{v_2c_2,v'c'}(\omega,T)
\end{align}
which is depicted in Fig.~\ref{fig:Dyson_Lambda}.

A detailed derivation of $\Xi^0$ is provided
  in supplemental materials~\cite{SM}.
Throughout this derivation, we assume the existence
  of a band gap that does not allow
  for a significant density of thermal carriers 
  at the temperature of interest,
  which is typically room temperature, where $E_g \gg k_B T$.

\begin{figure}[t]
  \includegraphics[width=0.8\linewidth]{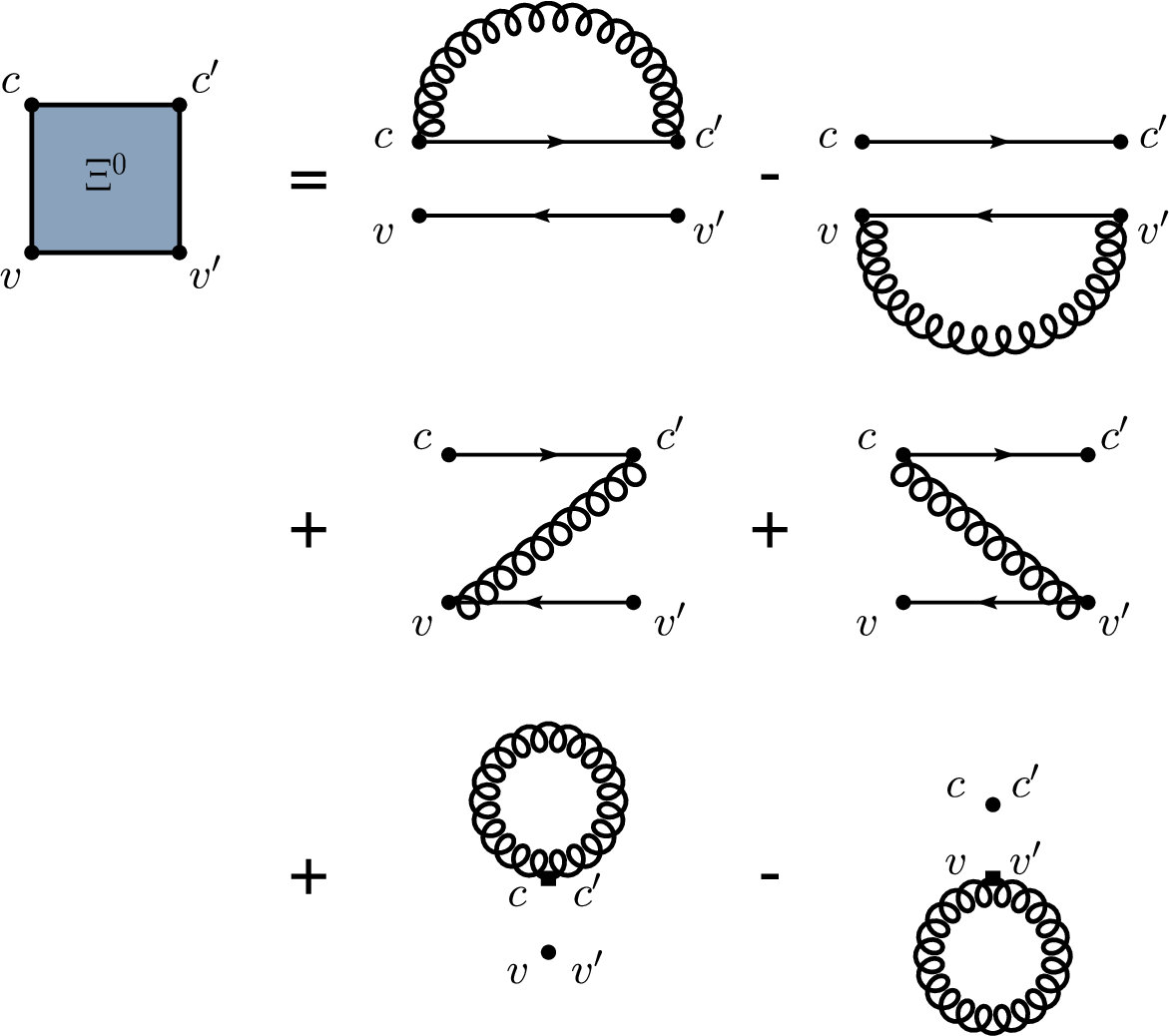}\\
  \caption{\label{fig:Xi0_all}
    The different contributions to the IEHPP self-energy $\Xi^0$,
    corresponding to a free electron-hole pair interacting with phonons.
    Contributions on the right-hand side of the equation are:
    the Fan-Migdal terms (first line),
    the phonon exchange terms (second line),
    and the Debye-Waller terms (third line).
    }
\end{figure}

The different contributions to the
  IEHPP
  self-energy
  are grouped in three terms, according to the topology
  of the diagrams depicted in Fig.~\ref{fig:Xi0_all}
\begin{align}
  \Xi^{0}_{vcv'c'}(\omega, T) =
    \Xi^{0^{\text{FM}}}_{vcv'c'}(\omega, T)
  + \Xi^{0^{\text{X}}}_{vcv'c'}(\omega, T)
  + \Xi^{0^{\text{DW}}}_{vcv'c'}(T) 
.
\end{align}
The Fan-Migdal term is further split in two contributions:
  the dynamic term (FMd) and the static term (FMs),
\begin{align}
  \Xi^{0^{\text{FM}}}_{vcv'c'}(\omega, T) =
    \Xi^{0^{\text{FMd}}}_{vcv'c'}(\omega, T)
  + \Xi^{0^{\text{FMs}}}_{vcv'c'}(T)
.
\end{align}

The dynamic FM term describes the propagation
  of an electron-hole pair being temporarily
  knocked into a different electron-hole state
  while absorbing/emitting a phonon.
Either the hole scatters into another valence state of different momentum,
  or the electron scatters into another conduction state.
Its diagrammatic expression
  is given by
\begin{align} \label{XiZeroDynFan}
  \Xi^{0^{\text{FMd}}}_{vcv'c'}& (i\omega_n) =
  - \frac{1}{\beta} \sum_m \sum_{\ipho} 
    \sum_{v''c''} D_\ipho(i\omega_m)
                  L^0_{v''c''v''c''}(i\omega_m + i\omega_n)\nonumber\\
  & %\qquad \qquad \qquad  
    \times \Big[  g_{v'v'' \ipho} g^{*}_{v v'' \ipho}
                   \delta_{cc''}\delta_{c''c'}
                + g_{cc'' \ipho} g^{*}_{c' c'' \ipho}
                   \delta_{vv''}\delta_{v''v'}
     \Big]
.
\end{align}
Performing the convolution of the
  bare electron-hole propagator $L^0$
  with the phonon propagator $D$, we obtain the analytic expression
\begin{align} \label{XiZeroDynFanAna}
 &  \Xi^{0^{\text{FMd}}}_{vcv'c'}(i\omega_n) = 
      \sum_{\ipho}
      \sum_{c''} g_{cc'' \ipho} g^{*}_{c' c'' \ipho} \delta_{vv'}\nonumber\\
     & \times
        \bigg[
        \frac{N_B(\omega_\ipho) - N_B(\eig_{c''} - \eig_{v})}
             {i\omega_n - (\eig_{c''} - \eig_{v}) + \omega_\ipho}
      + \frac{N_B(\omega_\ipho) +1 + N_B(\eig_{c''} - \eig_{v})}
             {i\omega_n - (\eig_{c''} - \eig_{v}) - \omega_\ipho}
        \bigg]\nonumber\\
  & + \sum_{v''} g_{v'v'' \ipho} g^{*}_{v v'' \ipho} \delta_{cc'}\nonumber\\
     & \times
        \bigg[
        \frac{N_B(\omega_\ipho) - N_B(\eig_c - \eig_{v''})}
             { i\omega_n - (\eig_c - \eig_{v''}) + \omega_\ipho}
      + \frac{N_B(\omega_\ipho) + 1 + N_B(\eig_c - \eig_{v''})}
             { i\omega_n - (\eig_c - \eig_{v''}) - \omega_\ipho}
        \bigg]
\end{align}
and one can safely assume that $N_B(\eig_c - \eig_{v},T) \ll N_B(\omega_\ipho,T)$.
The static FM term has the same topology
  as the dynamic FM term.
It is given by
\begin{align} \label{Xi0StatFan}
  \Xi^{0^{\text{FMs}}}_{vcv'c'} & = \frac{1}{\beta} \sum_m \sum_{\ipho}
    \sum_{v''c''} D_\ipho(i\omega_m) L^0_{v''c''v''c''}(i\omega_m)\nonumber\\
   & \times
          \Big[  g_{v'c'' \ipho} g^{*}_{v c'' \ipho}
                   \delta_{cc'}
                   \frac{1}{2} \big(\delta_{vv''} + \delta_{v''v'} \big)\nonumber\\
         &    \quad  + g_{cv'' \ipho} g^{*}_{c' v'' \ipho}
                   \delta_{vv'}
                   \frac{1}{2} \big(\delta_{cc''} + \delta_{c''c'} \big)
     \Big]
\end{align}
and its analytic expression is
\begin{align} \label{Xi0StatFanAna}
& \Xi^{0^{\text{FMs}}}_{vcv'c'} =  \sum_{\ipho}
      \delta_{vv'} \sum_{v''} g_{cv'' \ipho} g^{*}_{c' v'' \ipho} \nonumber\\
    & \times   \sym        \bigg[
            \frac{N_B(\omega_\ipho) + 1 + N_B(\eig_{c} - \eig_{v''})}
                 {(\eig_{c} - \eig_{v''}) + \omega_\ipho}
          + \frac{N_B(\omega_\ipho) - N_B(\eig_{c} - \eig_{v''})}
                 {(\eig_c - \eig_{v''}) - \omega_\ipho}
          \bigg]\nonumber\\
  & +\delta_{cc'} \sum_{c''} g_{v'c'' \ipho} g^{*}_{v c'' \ipho} \nonumber\\
  & \times \sym        \bigg[
            \frac{N_B(\omega_\ipho) + 1 + N_B(\eig_{c''} - \eig_v)}
                 {(\eig_{c''} - \eig_v) + \omega_\ipho}
          + \frac{N_B(\omega_\ipho) - N_B(\eig_{c''} - \eig_v)}
                 {(\eig_{c''} - \eig_{v}) - \omega_\ipho}
          \bigg]
\end{align}
where the symbol $\sym$ means that the terms in brackets
  should be symmetrized with the substitution $v,c \leftrightarrow v',c'$
  and a factor of $1/2$.
The static FM term includes transitions that cannot be described
  as an intermediate electron-hole pair,
  such as the hole being coupled to a conduction band state
  or the electron being coupled to a valence band state.
These transitions are only virtual, in the sense that they
  do not conserve energy and do not contribute
  to the imaginary part of the self-energy.

The next set of diagrams %contributing to $\Xi^0$
  are the phonon exchange (X) term,
  defined as
\begin{align}
 & \Xi^{0^{\text{X}}}_{vcv'c'}(i\omega_n)
  = \sum_{\ipho} g_{c'c \ipho} g^{*}_{v' v \ipho}\nonumber\\
 & \times   \frac{1}{\beta} \sum_m \big[ L^0_{vc',vc'}(i\omega_n + i\omega_m)
                               + L^0_{v'c,v'c}(i\omega_n + i\omega_m) \big]
                           D_{\ipho}(i\omega_m)
.
\end{align}
The analytic expression for this term is
\begin{align} \label{Xi0_X_final}
&  \Xi^{0^{\text{X}}}_{vcv'c'}(i\omega_n) = 
  - \sum_{\ipho} g_{cc' \ipho} g^{*}_{v v'\ipho}\bigg\{\nonumber\\
&   \bigg[
      \frac{ N_B(\omega_{\ipho}) - N_B(\eig_{c} - \eig_{v'})}
           {i\omega_n - (\eig_{c} - \eig_{v'}) + \omega_{\ipho}}
    + \frac{ N_B(\omega_{\ipho}) + 1 + N_B(\eig_{c} - \eig_{v'})}
           {i\omega_n - (\eig_{c} - \eig_{v'}) - \omega_{\ipho}}
    \bigg]\nonumber\\
  +
&   \bigg[
      \frac{ N_B(\omega_{\ipho}) - N_B(\eig_{c'} - \eig_{v})}
           {i\omega_n - (\eig_{c'} - \eig_{v}) + \omega_{\ipho}}
    + \frac{ N_B(\omega_{\ipho}) + 1 + N_B(\eig_{c'} - \eig_{v})}
           {i\omega_n - (\eig_{c'} - \eig_{v}) - \omega_{\ipho}}
    \bigg]
   \bigg\}
.
\end{align}
In this process, the electron emits a phonon
  that is being absorbed later on by the hole,
  or vice-versa.
This term is exclusively off-diagonal,
  since the electron and the hole exchange momentum
  and are being scattered into different states.
It will be non-zero
  only
  when ${\vk_{v'}=\vk_{v}+\vq_{\ipho}+\mv{G}}$
  and ${\vk_{c'}=\vk_{c}+\vq_{\ipho}+\mv{G}}$.

Finally, the Debye-Waller contribution to $\Xi^0$
  is simply the second-order interaction of the electron
  and the hole with the phonon modes, giving
\begin{align}
  \Xi^{0^{\text{DW}}}_{vcv'c'} = & \sum_{\ipho} \big(2N_B(\omega_\ipho) + 1\big)
    \Big[ g^{(2)}_{cc' \ipho \ipho } \delta_{vv'}
    - g^{(2)}_{v'v \ipho \ipho } \delta_{cc'} \Big]
.
\end{align}

Consider a non-interacting electron-hole excited state
  with energy $\eig_c - \eig_v$.
We can show that the diagonal component
  of the
  IEHPP
  self-energy for this state is
\begin{align}
 \Xi^{0}_{vcvc} (\eig_c - \eig_v) = &
   \Sigma_{cc}(\eig_c) - \Sigma_{vv}(\eig_v)
.
\end{align}
This is the expected result.
Without the electron-hole interaction,
  the corrections to the optical excitations
  are simply given by the
  electron-phonon interaction 
  corrections to the one-particle energies
  that make up the transitions.
Since the imaginary part of the self-energy
  has opposite signs for electrons and holes,
  we also have that
\begin{align}
 \vert \Im \Xi^{0}_{vcvc} (\eig_c - \eig_v) \vert = &
   \vert \Im \Sigma_{cc}(\eig_c) \vert + \vert \Im \Sigma_{vv}(\eig_v) \vert
.
\end{align}
The broadening of a non-interacting electron-hole transition
  is thus the sum of the broadenings of the one-particle states.

\subsection{Exciton-phonon self-energy}

The exciton-phonon self-energy $\Xi$ is obtained by expanding
  the interacting exciton propagator $\Lambda$ in the
  bare exciton basis as
\begin{align} \label{Lambdaexpandtwo}
  \Lambda_{SS'}(\tau)  = & -\sum_{n=0}^{\infty} (-1)^n
    \int_{0}^{\beta}d\tau_1 \dots \int_{0}^{\beta}d\tau_n\nonumber\\
  & \times
    \braket{\Ttau c_{S}(\tau) c^{\dagger}_{S'}(0)
      \hat V_{ep}(\tau_1)\dots \hat V_{ep}(\tau_n)}_0 \nonumber\\
  & + \Lambda^{\text{non-excitonic}}_{SS'}(\tau)
\end{align}
where the last term regroups all the "non-excitonic" diagrams
  and will be discussed further below.
In this formulation,
  the electron-hole interaction is already included in $L$
  (the bare exciton propagator without considering coupling of the exciton to phonons),
  and one needs to expand the perturbation in powers of $V_{ep}$ only.
The solution is thus analogous to the one-particle case,
  i.e., considering an exciton as a single particle interacting with the phonons.
The physical motivation for this formulation is that
  the two kind of processes
  occur on different time scales.
The electron-hole interaction is much faster
  than the electron-phonon interaction,
  since the plasmon frequency
  is much larger than the typical phonon frequency.

We define the exciton annihilation operator as
\begin{equation} \label{eq:cSdef}
  c_S = \sum_{vc} A^{S*}_{vc} c^{\dagger}_v c_c
\end{equation}
with the coefficient $A^S$ from solving the BSE, 
and we treat these excitations as bosons.
This treatment is not exact,
  since the Bose commutation relations are not exactly fulfilled
  by the operators in Eq.~\eqref{eq:cSdef}.
However, in the low exciton density limit \cite{laikhtman_are_2007}
  we can take
\begin{equation}
  \big[c_S, c^{\dagger}_{S'} \big]
  = c_S c^{\dagger}_{S'} - c^{\dagger}_{S'} c_S
  = \delta_{SS'}
.
\end{equation}
The bare exciton propagator
  is diagonal in the exciton basis and is formally defined as
\begin{equation}
  L_{SS'}(\tau) = - \braket{c_S(\tau) c^{\dagger}_{S'}(0)}_{2p 0}
\end{equation}
where the subscript $2p 0$ refers to the ground state of a Hamiltonian
  that includes electron-hole interactions through the BSE kernel
  but does not include electron-phonon interactions.
In this picture, the time
  dependence of the operators is given by
\begin{equation}
  c_{S}(\tau) = e^{\tau \hat H^{2p}} c_{S}(0) e^{-\tau \hat H^{2p}}
\end{equation}
with the two-particle Hamiltonian
  ${\hat H^{2p} = \sum_S \Omega_S c^{\dagger}_S c_S}$.
The perturbation terms $V_{ep} = V^{(1)}_{ep} + V^{(2)}_{ep}$ appearing
  in Eq.~\eqref{Lambdaexpandtwo}
  may be expressed in the exciton basis as
\begin{align} \label{VoneSS}
  V^{(1)}_{ep} = \sum_{SS' \ipho} g_{SS' \ipho}
                 A_{\ipho} c^{\dagger}_{S} c_{S'}
\end{align}
and
\begin{align} \label{VtwoSS}
  V^{(2)}_{ep} = \sum_{SS' \ipho \ipho'} g^{(2)}_{SS' \ipho \ipho'}
                 A_{\ipho} A_{\ipho'} c^{\dagger}_{S} c_{S'}
\end{align}
Here, as defined above, $A_{\ipho}$ is a sum of a phonon creation and annihilation operator,
  (not to be confused with the electron-hole coefficients $A^{S}_{vc}$).
The first- and second-order exciton-phonon coupling matrix elements are
  the transition amplitude between two exciton states
  mediated by the electron-phonon coupling potential, e.g.
\begin{align} \label{gsspre}
    g_{S S' \ipho}
    = \braket{S \vert \Big( \sum_{\ieig\jeig} g_{\ieig\jeig\ipho} c_{\ieig}^\dagger c_{\jeig} \Big) \vert S'}
    = \sum_{\ieig\jeig} g_{\ieig\jeig\ipho}
      \braket{c_{S} c_{\ieig}^\dagger c_{\jeig} c_{S'}^{\dagger}}
.
\end{align}
Using Eq.~\eqref{eq:cSdef} and~\eqref{eq:commut}
  to simplify the expectation values % in Eq.~\eqref{gsspre},
  and discarding terms like
  $\delta_{SS'}\sum g_{vv\ipho}$
  which produce disconnected diagrams, % associated with the vacuum amplitude,
  we arrive at the expressions

\begin{align} \label{gonetwop}
    g_{S S' \ipho} = \sum_{vc,v'c'} A_{vc}^{S*} A_{v'c'}^{S'}
                \Big[ g_{cc' \ipho} \delta_{vv'}
                    - g_{v'v \ipho} \delta_{cc'} \Big]
\end{align}
and
\begin{align}
  g^{(2)}_{SS' \ipho\ipho} =  \sum_{vc,v'c'} A_{vc}^{S*} A_{v'c'}^{S'}
                \Big[ g^{(2)}_{cc' \ipho \ipho} \delta_{vv'}
                    - g^{(2)}_{v'v \ipho \ipho} \delta_{cc'} \Big]
.
\end{align}
Momentum conservation dictates that $\vq_{S}=\vq_{S'}+\vq_{\ipho}$
  for $g_{SS' \ipho}$ to be nonzero,
  while $\vq_{S'}=\vq_{S}$ for $g^{(2)}_{SS' \ipho\ipho}$.
It is worth writing explicitly all the wavevectors
  in these expressions.
For an exciton $S'$ with
  center-of-mass
  momentum $\vq_{S'}=\vQ$ %and band index $n_{S'}=S'$,
  and an exciton $S$ with
  center-of-mass
  momentum $\vq_{S}=\vQ+\vq$,
  the first-order matrix element writes as
\begin{align}
  g_{S S' \ipho}(\vQ, \vq)
    = & \sum_{vcc' \vk} A^{S\vQ+\vq *}_{vc\vk} A^{S'\vQ}_{vc'\vk}
                      g_{cc' \ipho}(\vk+\vQ, \vq)\nonumber\\
    - & \sum_{vv'c \vk} A^{S\vQ+\vq *}_{vc\vk} A^{S'\vQ}_{v'c\vk+\vq}
                      g_{v'v \ipho}(\vk, \vq)
\end{align}
where $A^{S\vQ}_{vc\vk}$  is the coefficient
  of the exciton envelope function in k-space 
  for a hole at $\vk$
  and an electron at $\vk+\vQ$.
We note that $(\vQ,\vq)$ are not Fourier components of the matrix element;
  they are momentum information in fact that is contained
  in the quantum numbers $S$ and $S'$.
Here we use a somewhat redundant notations to draw out the physics
  of center-of-mass momentum conservation of the excitons.
The second-order matrix elements between two exciton states
  with the same momentum $\vQ$ are
\begin{align}
  g^{(2)}_{SS' \ipho\ipho}(\vQ,\vq) & =
      \sum_{vcv'c'\vk}
      A_{vc\vk}^{S\vQ*} A_{v'c'\vk}^{S'\vQ}
\nonumber\\
      & \times \Big[ g^{(2)}_{cc' \ipho \ipho}(\vk, \vq,-\vq) \delta_{vv'}
                    - g^{(2)}_{v'v \ipho \ipho}(\vk, \vq, -\vq) \delta_{cc'} \Big]
.
\end{align}
Note that the second-order matrix elements do not transfer momentum between the exciton states, even when $\vq\neq 0$.

\begin{figure}[t]
  \includegraphics[width=0.8\linewidth]{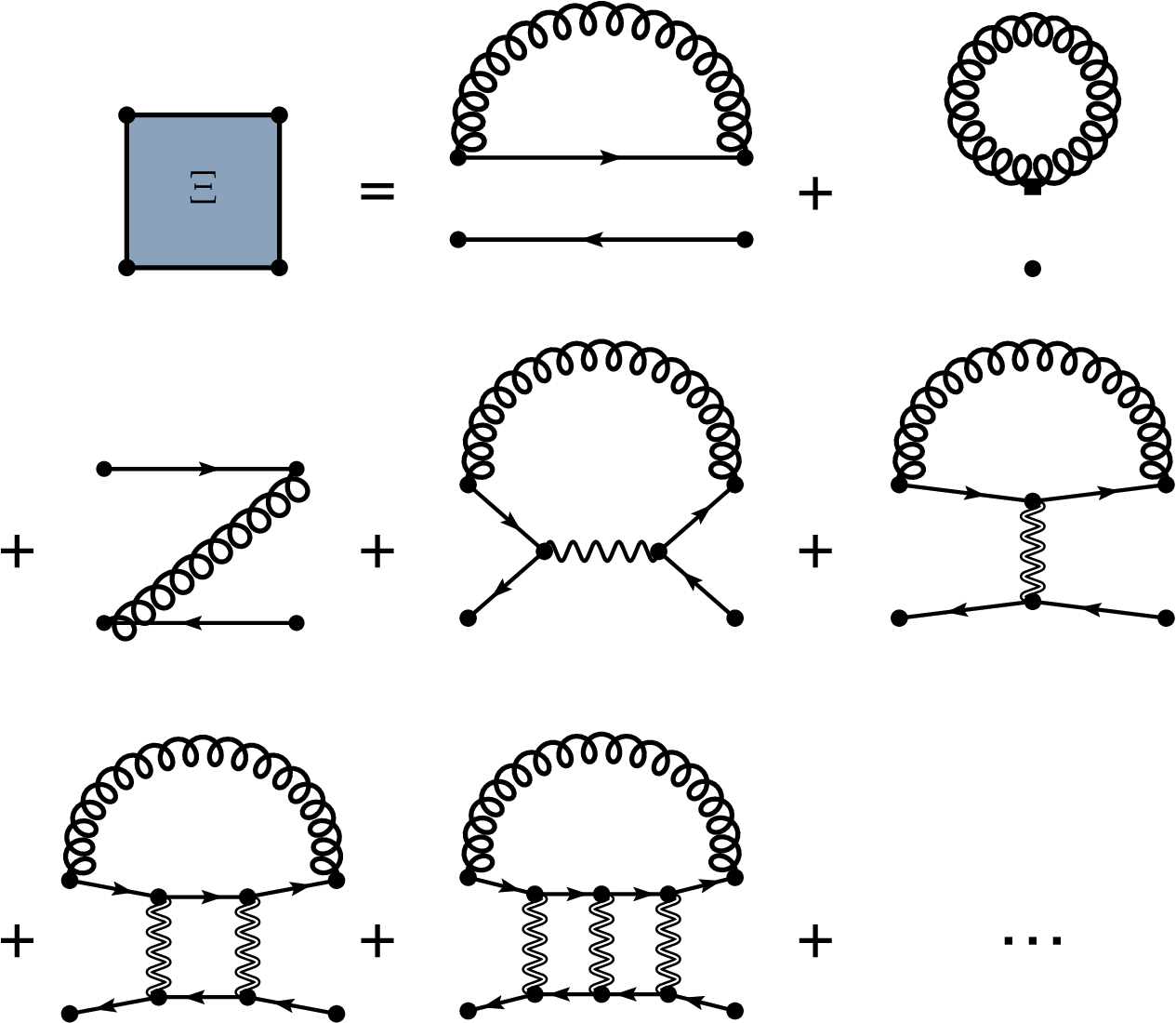}
  \caption{\label{fig:Xi_many}
    Some of the diagrams contributing to the exciton-phonon self-energy ($\Xi$)
    in the quasiparticle basis.
    This series contains all the diagrams
    with one or many Coulomb interaction lines (wavy lines)
    between the phonon vertices,
    in addition to the diagrams of
    the IEHPP self-energy
    ($\Xi^0$).
    Diagrams consisting of a phonon exchange line and Coulomb interaction lines
    are present in this series, but are not depicted.
    }
\end{figure}

We may now compute the self-energy %in the same way as in the one-particle case
  from Eq.~\eqref{Lambdaexpandtwo}.
Some of the diagrams using quasiparticle (electron/hole) basis
  for the exciton-phonon self-energy up to second order
  in electron-phonon coupling are illustrated in Fig.~\ref{fig:Xi_many}. 
The Debye-Waller diagram coming from the second-order matrix element is
\begin{equation}
  \Xi^{\text{DW}}_{SS'} = \sum_{\ipho} g^{(2)}_{SS' \ipho \ipho}
                                \big[2N_B(\omega_\ipho) + 1\big]
.
\end{equation}
This is the same result as for the case without
  electron-hole interactions, meaning that,
  using the basis transformation rule of Eq.~\eqref{eq:basistransform}, we find
  $\Xi^{\text{DW}}_{SS'} = \Xi^{0^{\text{DW}}}_{SS'}$.

The first-order coupling matrix elements yield the Fan-Migdal term. 
In the perturbative expansion of $\Lambda$,
  the first term of Eq.~\eqref{Lambdaexpandtwo}
  produces only the dynamic part of the FM term,
  since only this part can be expressed in terms of $g_{SS'\ipho}$.
It is given by
\begin{align}
  \Xi^{\text{FMd}}_{SS'} (i\omega_n)
  = &
  -\!\frac{1}{\beta}\!%
  \sum_{m \ipho S''}
        g_{SS'' \ipho} g^{*}_{S' S'' \ipho}
        D_\ipho(i\omega_m) L_{S''S''}(i\omega_m\!+\!i\omega_n)
\end{align}
with the analytic expression
\begin{align} \label{XiDynFan}
&  \Xi^{\text{FMd}}_{SS'} (i\omega_n)
  =  \sum_{\ipho} \sum_{S''}  g_{SS'' \ipho} g^{*}_{S' S'' \ipho}\nonumber\\
  & \times    \bigg[
        \frac{N_B(\omega_\ipho) - N_B(\Omega_{S''})}
             {i\omega_n - \Omega_{S''} + \omega_\ipho}
      + \frac{N_B(\omega_\ipho) + 1 + N_B(\Omega_{S''})}
             {i\omega_n - \Omega_{S''} - \omega_\ipho}
      \bigg]
.
\end{align}
We can verify that in the limit where the electron-hole interaction vanishes,
  that is, when each exciton is composed of a single electronic transition,
  Eq.~\eqref{XiDynFan} reduces to Eq.~\eqref{XiZeroDynFanAna}.

In the exciton basis, the phonon exchange term
  doesn't need to be added;
  it is included in the dynamic FM term.
This term arises from the cross terms between valence-to-valence
  and conduction-to-conduction bands coupling
  when squaring Eq.~\eqref{gonetwop}.
Since the exciton states mix several electronic transitions
  across the Brillouin zone,
  the phonon exchange term does contribute to the diagonal elements
  of $\Xi$.

All other diagrams of $\Lambda$ in which the scattered state
  cannot be projected onto the exciton basis
  are contained in $\Lambda^{\text{non-excitonic}}$.
One such diagram is obtained from the static FM term.
Therefore, we complete the interacting FM self-energy
  by adding the static FM term of Eq.\eqref{Xi0StatFanAna} so that
\begin{align}
  \Xi^{\text{FM}}_{SS'} (i\omega_n) = \Xi^{\text{FMd}}_{SS'} (i\omega_n)
                              + \Xi^{\text{FMs}}_{SS'}
\end{align}
  with $\Xi^{\text{FMs}}_{SS'} = \Xi^{0^{\text{FMs}}}_{SS'}$.
Using the non interacting static FM term here is an approximation.
Since the static FM term is composed
  of virtual transitions across the band gap,
  the relative error induced by this approximation
  is at most on the order of $E^S_b/E_g$,
  where $E^S_b$ is the binding energy of the exciton.

As formulated, in practical implementation, there will be an
  additional term in the self-energy,
  which contributes
  to $\Lambda^{\text{non-excitonic}}$.
Most of the time,
  the basis of valence and conduction bands used to expand
  the exciton wavefunctions is not complete;
  only a subset of the bands are used.
Moreover, one typically doesn't compute all the possible
  excitons states, but only the first few solutions of the BSE
  with the lowest eigenvalues.
Therefore, there will be a missing contribution to $\Xi^{\text{FMd}}$,
  which we call the completion term, written as $\Xi^{\text{C}}$.
Since the missing contribution involves high-energy states,
  this term can be computed by
  taking these states as free electron-hole pairs.
One can construct $\Xi^{\text{C}}$
  from the expression of $\Xi^0$,
  where the intermediate states $v''$ and $c''$
  are projected outside of the basis of excitons that were computed.
An example of this procedure is given in Sec.~\ref{sec:Application}.

Collecting all terms, the final expression for the exciton-phonon
  self-energy reads
\begin{align} \label{XiFinal}
  \Xi_{SS'} (\omega,T)\!=\!\Xi^{\text{FMd}}_{SS'}(\omega,T)%
                      \!+\!\Xi^{\text{FMs}}_{SS'}(T)%
                      \!+\!\Xi^{\text{DW}}_{SS'}(T)%
                      \!+\!\Xi^{\text{C}}_{SS'}(\omega,T)%
.
\end{align}
Thus, $\Xi$ contains all the diagrams of $\Xi^0$
  but with addition of the electron-hole interaction diagrams,
  as shown in Fig.~\ref{fig:Xi_many}.
Note that $\Xi^{\text{C}}_{SS'}$ is complex and frequency-dependent,
  but its imaginary part is non-zero only at frequencies
  far from the bare exciton energy,
  and its real part varies smoothly near the exciton energy.
Hence, near the exciton energy ($\Omega_S$) we may consider that
  only the dynamic Fan-Migdal term contributes to the imaginary part
  of the self-energy, which is
\begin{align} \label{ImXi}
 & \Im \Xi_{SS'} (\omega,T) =
   - \pi \sum_{\ipho} \sum_{S''}  g_{SS'' \ipho} g^{*}_{S' S'' \ipho}
\nonumber\\
  & \times  \Big[ 
        \big( N_B(\omega_\ipho, T) - N_B(\Omega_{S''}, T) \big)
          \delta(\omega - \Omega_{S''} + \omega_\ipho) 
\nonumber\\
  &   +  \big(N_B(\omega_\ipho, T) + 1 + N_B(\Omega_{S''}, T) \big)
          \delta(\omega - \Omega_{S''} - \omega_\ipho)
      \Big]
.
\end{align}
The temperature renormalization of the exciton energies and their lifetime
  can finally be computed according to Eqs.~\eqref{OmegaRen} and~\eqref{OmegaBrd}.

\subsection{Approximate expressions}

Since the excitation energies
  are usually much larger than the phonon frequencies,
  it is a safe approximation
  to use the fact that $N_B(\Omega_S,T) \ll N_B(\omega_\ipho,T)$
  and write for the diagonal part of the self energy
\begin{align}
&  \Xi^{\text{FMd}}_{SS} (\omega, T) =
    \sum_{\ipho} \sum_{S''}
      \vert g_{SS'' \ipho} \vert^2\nonumber\\
& \times \bigg[
        \frac{N_B(\omega_\ipho, T)}
             {\omega - \Omega_{S''} + \omega_\ipho + i\eta}
      + \frac{N_B(\omega_\ipho, T) + 1}
             {\omega - \Omega_{S''} - \omega_\ipho + i\eta}
      \bigg]
.
\end{align}

Let us compare the full exciton-phonon self-energy with
  approximate expressions found in literature.
We first define
  the "uncorrelated exciton" (UE) approximation
  to the exciton-phonon self-energy:
\begin{align} \label{XiFXA}
  \Xi_{SS}^{\text{UE}} = & \sum_{vc} \vert A^S_{vc} \vert^2
    \Xi^0_{vcvc}(\eig_c - \eig_v)
    =  \sum_{vc} \vert A^S_{vc} \vert^2 
    \Big[ \Sigma_c(\eig_c) - \Sigma_v(\eig_v) \Big]
\end{align}
This approach is equivalent to the one used in previous studies~%
  \cite{Marini2008,qiu_optical_2013,molina-sanchez_temperature-dependent_2016}.
Several approximations have been made between Eqs.~\eqref{XiFinal}
  and \eqref{XiFXA}.
First, the
  UE
  self-energy is constructed using the
  independent electron-hole-pair-phonon propagator (IEHPP).
Then, only the diagonal elements of $\Xi^0$ are
  used when transforming from the one-particle basis to the exciton basis. 
Finally, the self-energy is being evaluated at the non interacting
  transition energies.
The
  uncorrelated exciton approximation
  reduces tremendously the computational effort
  needed to obtain the self-energy, since the knowledge of the finite-momentum
  exciton states is no longer required. 
However, it does not yield accurate results for the lifetime
  of bound excitons, as we show in the next section.

In this approximation,
  the uncorrelated exciton is just a wavepacket
  composed of electron-hole pairs that do not interact with each other
  and hence there is no binding energy.
This becomes apparent when looking at the imaginary part of the self-energy.
Using Eq.~\eqref{XiZeroDynFanAna},
  the imaginary part of $\Xi^{\text{UE}}_{S}$
  writes as
\begin{align} \label{ImXiZero}
  \Im \Xi^{\text{UE}}_{SS} = &
     - \pi \sum_{\ipho} \sum_{vc} \vert A^S_{vc} \vert^2 \bigg\{\nonumber\\
&   \sum_{c''} \vert g_{cc'' \ipho} \vert^2
 \Big[ \big(N_B(\omega_\ipho) + f(\eig_{c''})\big)
            \delta(\eig_c - \eig_{c''} + \omega_\ipho)\nonumber\\
& \quad        + \big(N_B(\omega_\ipho) + 1 - f(\eig_{c''})\big)
            \delta(\eig_c - \eig_{c''} - \omega_\ipho)
      \Big]\nonumber\\
  + & \sum_{v''} \vert g_{vv'' \ipho} \vert^2
      \Big[ \big(N_B(\omega_\ipho) + f(\eig_{v''})\big)
            \delta(\eig_v - \eig_{v''} + \omega_\ipho)\nonumber\\
& \quad        + \big(N_B(\omega_\ipho) + 1 - f(\eig_{v''})\big)
            \delta(\eig_v - \eig_{v''} - \omega_\ipho)
      \Big] \bigg\}
\end{align}
Compare this expression with
  Eq.~\eqref{ImXi},
  where the scattering occurs between exciton states.
This very drastic approximation in past studies
  creates a qualitative difference
  in the physics of the exciton 
  in the low temperature limit,
  where ${N_B(\omega_\ipho,T) \rightarrow 0}$.
For the lowest bound exciton, from Eq.~\eqref{ImXi},
  no states are available for scattering through
  phonon emission.
Hence, the electron-phonon interaction does not contribute
  to its inverse lifetime.
The physical lifetime of this exciton comes from other processes,
  such as radiative recombination or scattering by impurities
  \cite{spataru_theory_2005-1}.
In the UE approximation however, scattering events still occur
  because the binding energy of the initial and final states are ignored,
  and the lowest bound exciton has an unphysical lifetime at zero temperature.

For the temperature-dependent energy shift of the excitonic peaks,
  the accuracy of the UE approximation is unknown at this point.
In the following section, we gain some intuition on this matter
  by computing the self-energy in a model system.

\section{Application to a model system}
\label{sec:Application}

To illustrate the theory, we present
  a two-band model in two dimensions,
  and compute the energy and lifetime
  of optical excitations as a function of temperature.
We use the triangular lattice exciton model
  proposed by D. Gunlycke and F. Tseng~\cite{gunlycke_triangular_2016},
  which mimics the spin-dependent band dispersion of transition metal dichalcogenides
  near the $K$ and $K'$ valleys,
  and yields realistic exciton binding energies.
The main equations are reported below,
  and we refer the reader to Ref.~\onlinecite{gunlycke_triangular_2016}
  for further details of this model.

\subsection{One-particle and two-particle Hamiltonians}

\begin{figure}[t]
  \includegraphics[width=\linewidth]{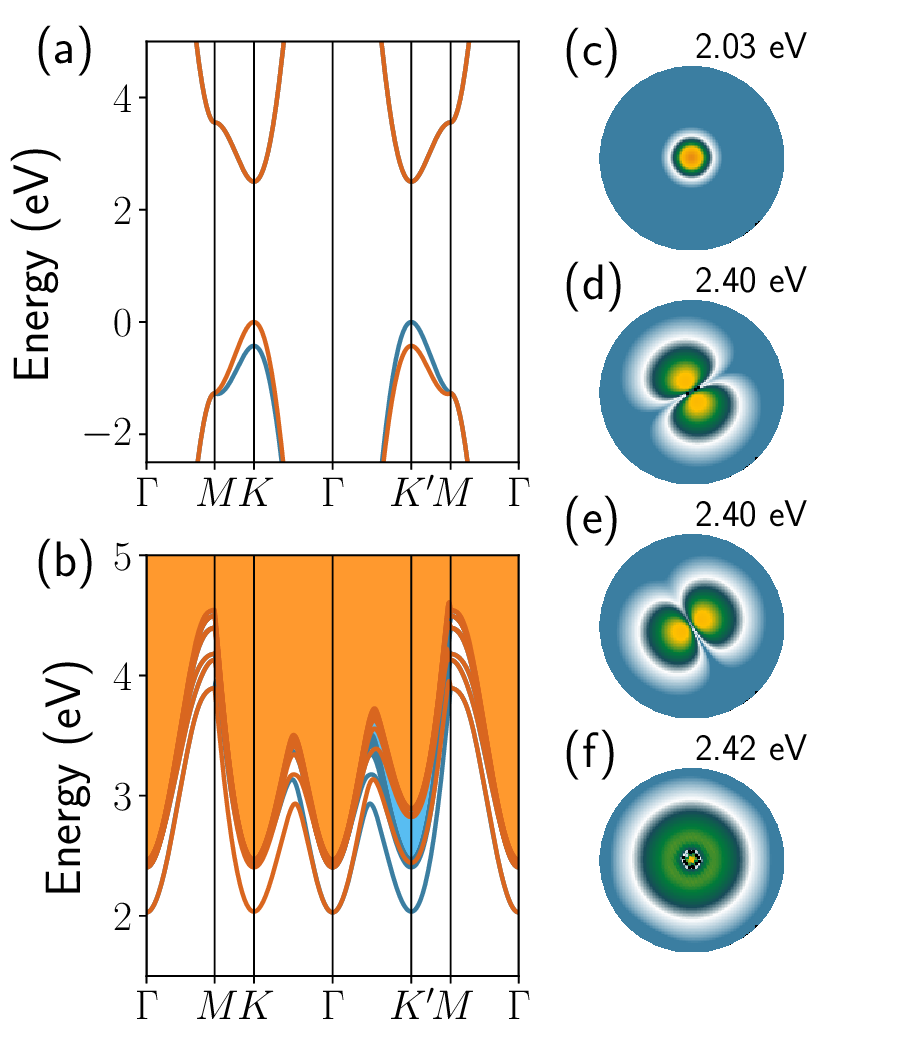}
  \caption{\label{fig:twobands} %{\color{red} plot-340-107-tb-multiple.png}
  \textbf{a\textendash}%
  Two-band model electronic structure
  for the different spin states $\ket{\uparrow}$ (orange)
  and $\ket{\downarrow}$ (blue).
  \textbf{b\textendash}%
  Exciton band structure
  for the different spin states
  $\ket{\uparrow\downarrow}$
  (orange)
  and
  $\ket{\downarrow\uparrow}$
  (blue).
  The filled regions represent the continuum of excited states.
  \textbf{c-f\textendash}%
  Real space wavefunction $\vert A^S(\vR, \vQ)\vert^2$
  of the first four optically accessible excitons
  ($S\!=\!1-4, \vQ\!=\!\Gamma$).
  }
\end{figure}

The one-particle tight-binding Hamiltonian is
  $H=\sum_{n\ispin} H_{n\ispin}$ with
\begin{equation} \label{eq:1pH}
  H_{n\ispin} =
    \sum_{\vR, \boldsymbol{\delta}} t_{n\ispin\boldsymbol{\delta}}
      c^{\dagger}_{n\ispin \vR + \boldsymbol{\delta}}  c_{n\ispin \vR}
  + \sum_{\vR} \eig_{n} c^{\dagger}_{n\ispin \vR}  c_{n\ispin \vR} 
\end{equation}
  where $c^{\dagger}_{n\ispin \vR}$ and $c_{n\ispin \vR}$
  are creation and destruction operators for an electron
  in the~ $n^{\text{th}}$ orbital on the lattice site~$\vR$
  with spin~$\ispin$.

The first term describes the hopping between
  a site and one of its six closest neighbours,
  where $\boldsymbol{\delta}$ is the lattice vector 
  from one site to the other.
The hopping amplitude is composed of a spatial amplitude $t_n$
  and a spin-orbit coupling parameter $\tilde t_{n}$ according to
\begin{equation}
  t_{n\ispin\boldsymbol{\delta}} = t_{n} + 4 i \sigma \tilde t_{n} \sin(\mv K \cdot \boldsymbol{\delta})
\end{equation}
  where $\sigma=\pm 1/2$ and
  $\mv K$ is one corner of the 2D
  hexagonal
  Brillouin zone.
The second term of Eq.~\eqref{eq:1pH} is diagonal in lattice site,
  with $\eig_{n}$ being the on-site parameters.
The solutions for the one-particle energies
  at wavevector $\vk$ are
\begin{equation}
  \eig_{n\sigma\vk} = \eig_n + \sum_{\boldsymbol{\delta}} t_{n\sigma\boldsymbol{\delta}} e^{-i\vk\cdot\boldsymbol{\delta}}
.
\end{equation}

In order for this Hamiltonian to reproduce the main features of
  typical transition metal dichalcogenides
  for the valence (v) and conduction (c) band,
  the parameters are chosen by imposing that
  the band structure has a band gap $E_g$ located
  at the $K$ and $K'$ valleys,
  with band effective masses $m^*$
  and a splitting $\Delta$ between different spin bands.
The on-site and hopping parameters are thus 
\begin{align*}
  \begin{array}{lll}
   \eig_c = 3 t + E_g      \quad & t_c = t  \quad & \tilde t_c = 0 \\
   \eig_v =-3 t - \Delta/2 \quad & t_v = -t \quad & \tilde t_v = \Delta/18 \\
  \end{array}
\end{align*}
with $t = {2 \hbar^2}/{3m^* a^2}$ and $a$ is the lattice constant.
The resulting band structure is shown in Fig.~\ref{fig:twobands}(a)
  with the parameters
  $m^*=0.49$, $a=3.13$~Bohr, $E_g=2.5$~eV, and $\Delta=425$~meV.
Each spin channel admits one valence band and one conduction band.
The two spin channels are degenerate for the conduction band,
  while for the valence band, the splitting of the up and down spins
  due to spin-orbit coupling reaches a maximum at the $K$ and $K'$ points
  in the Brillouin zone.

The BSE Hamiltonian, which yields the exciton bands,
  is made of two terms: the kinetic energy,
  and the Coulomb interaction
  between the electron and the hole.
Translational symmetry implies that the exciton states %,
  possess a well-defined center-of-mass momentum $\vQ$,
  and the BSE Hamiltonian writes
\begin{equation}
  H_{\ispin \vQ} = 
    \sum_{\vR, \boldsymbol{\delta}} T_{\ispin\vQ \boldsymbol{\delta}}
      b^{\dagger}_{\ispin \vR + \boldsymbol{\delta}}  b_{\ispin \vR}
  + \sum_{\vR} \Big( \eig_c - \eig_v - V(\vR) \Big)
    b^{\dagger}_{\ispin \vR}  b_{\ispin \vR} 
\end{equation}
where $b^{\dagger}_{\ispin \vR}$ creates an electron at $\vR_e=\vR$
  with spin $\ispin$, and a hole at $\vR_h = 0$
  with spin $-\ispin$,
  meaning that no spin-flip occurs in this process.
The BSE hopping parameter carries the information on the exciton momentum,
  and writes
\begin{align}
  T_{\ispin\vQ \boldsymbol{\delta}} & = t_{c\sigma\boldsymbol{\delta}}e^{-i\vQ \cdot \boldsymbol{\delta} / 2}
                         - t_{v\sigma\boldsymbol{\delta}}e^{ i\vQ \cdot \boldsymbol{\delta} / 2}
.
\end{align}
Together, the on-site terms that involve $\eig_c$ and $\eig_v$
  and the hopping term form the kinetic energy component
  of the excitons.
In this model,
  the Coulomb interaction term
  is simply taken to be
\begin{align}
  V(\vR) =
    \begin{cases}
      \Delta v_0 \quad & (\vR=0)\\
      \frac{e^2}{4\pi\epsilon R} \quad & (\vR \neq 0)\\
    \end{cases}
\end{align}
where $\epsilon$ is the dielectric constant of the medium
  and $\Delta v_0$ is the difference between the screened Coulomb interaction
  and the bare exchange integral at $\vR=0$,
  which we set to $\Delta v_0=1.6$~eV.
Note that, for this model, the spin degrees of freedom do not mix,
  and at $\Gamma$,
  both spin channels are equivalent
  and can be interpreted as singlet exciton states.
In what follows,
  we will assume that the phonons are non-magnetic
  and cannot flip the spin.
The index $\sigma$ will thus be omitted.

Due to the sparse nature of a Hamiltonian with nearest-neighbour hopping,
  the BSE can be solved efficiently in real space for all $\vQ$-vectors.
Solving the BSE yields the exciton energies $\Omega_{\vQ S}$
  and wavefunctions in real space $A^S(\vR, \vQ)$.
Their Fourier transform give the electron-hole coefficients
  $A^S(\vk, \vQ)$.
Note that, within this model,
  it is not necessary to specify the one-particle band indices
  for these coefficients, since there is only a single valence band
  and conduction band.

The exciton band structure
  is shown in Fig~\ref{fig:twobands}(b).
The binding energy of the first four optical excitons
  is between 100 and 500~meV,
  and their wavefunction in real space is depicted in Fig.~\ref{fig:twobands}(c-f).

\subsection{Exciton-phonon coupling}

\begin{figure}[t]
  \includegraphics[width=\linewidth]{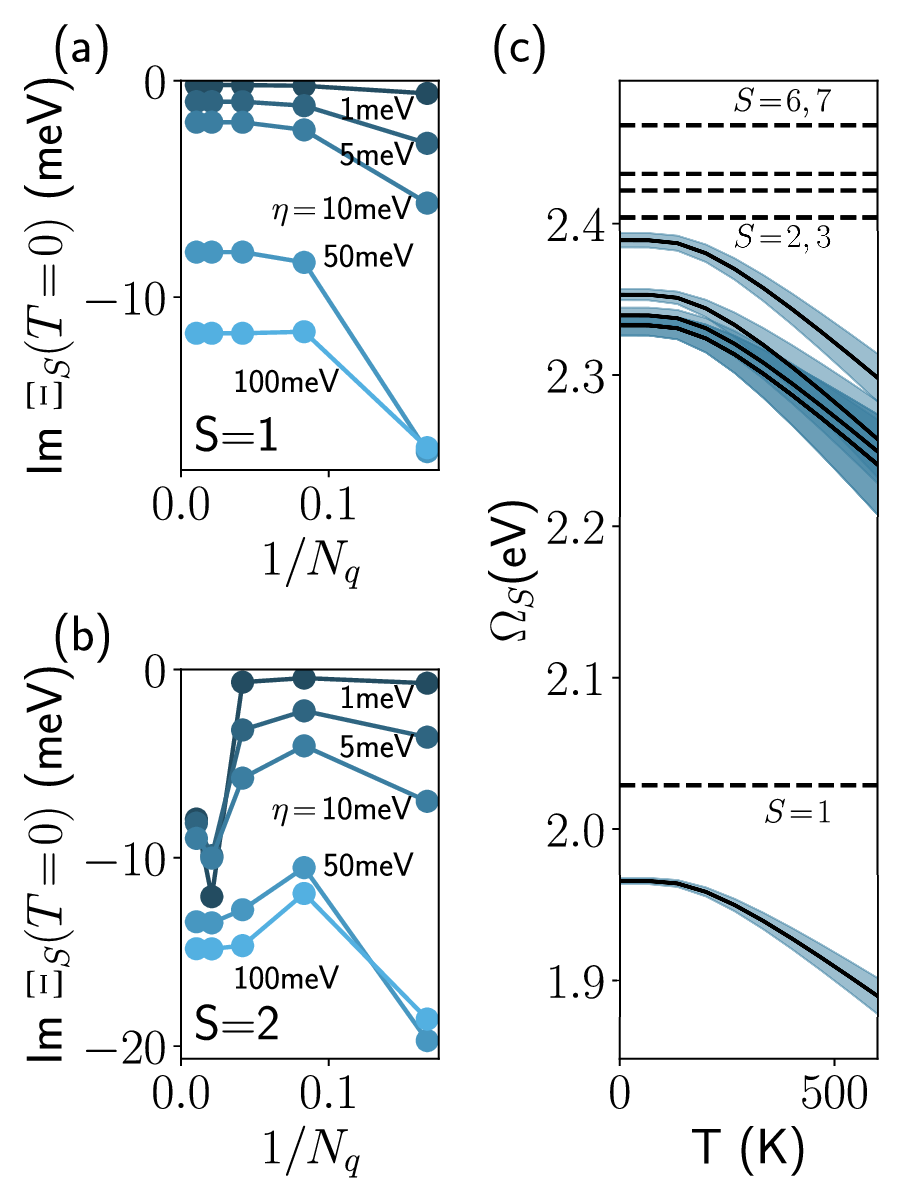}
  \caption{\label{fig:OmegaT}
    \textbf{a-b\textendash}%
    Convergence of the
    imaginary part of the self-energy
    with respect to the number of $\vq$-points
    for the lowest two excitons ($S\!=\!1,2$).
    $\vq$-grids are increasing (right to left)
    from $6 \times 6$ to $96 \times 96$,
    for different broadening parameter $\eta$.
    \textbf{c\textendash}%
    Temperature dependence of the optical excitation energies (solid line)
    and their line broadening (filled color).
    The dash lines indicate the bare exciton energies,
    without electron-phonon coupling.
}
\end{figure}

We model the lattice vibration spectrum
  as a single dispersionless phonon band with frequency $\omega_0$.
The electron-phonon coupling strength is also taken to be independent
  of the phonon wavevector,
  and we use a single parameter $g$ to represent
  intra-band coupling ($g_{cc'}\!=\!g_{vv'}\!=\!g$ ; $g_{cv}\!=\!0$).
For simplicity of notation,
  we write the intra-band coupling
  constants as $g_c$ and $g_v$.
In the present calculation, we set $\omega_0\!=\!50$~meV and $g\!=\!250$~meV.

The IEHPP
    self-energy for a hole at $\vk$ and an electron at $\vk + \Gamma$
    is diagonal in $\vk$ indices and is given by
\begin{align} \label{eq:IEHPPtwobandk}
&   \Xi^{0}_{\Gamma \vk}(\omega, T)
      = \sum_{\vq} \Xi^{0}_{\Gamma \vk, \vq}(\omega, T)
      = \sum_{\vq}
\\
&       \bigg[
        \frac{ \vert g_c \vert^2 P_{\pm}(T)}
             {\omega - (\eig_{\vk+\vq c} - \eig_{\vk v}) \pm \omega_0 + i \eta}
     +  
        \frac{ \vert g_v \vert^2 P_{\pm}(T)}
             {\omega - (\eig_{\vk c} - \eig_{\vk-\vq v}) \pm \omega_0 + i \eta}
        \bigg]
\nonumber
\end{align}
where $P_{+}(T) =  N_{B}(\omega_0, T) $ and $P_{-}(T) = N_B(\omega_0, T) + 1$
  correspond to phonon absorption and emission channels, respectively,
  and both channels must be added.
In the exciton basis,
  $\Xi^{0}$ is non-diagonal
  in the indices $S,S'$,
  but remains diagonal in the wavevector $\vQ$.
Here, we consider only the optical excitons,
  and we use $\Xi_{\Gamma S}$ to denote the diagonal
  elements of $\Xi$.
The IEHPP
    self-energy in the exciton basis is
\begin{align} \label{eq:IEHPPtwobandS}
    \Xi^{0}_{\Gamma S}(\omega, T)
    & = \sum_{\vq} \Xi^0_{\Gamma S, \vq}(\omega, T)
      = \sum_{\vq} \sum_{\vk} \vert A^S(\vk, \Gamma) \vert^2 \Xi^0_{\Gamma \vk,\vq}(\omega, T)
\end{align}
and we have introduced the symbols $\Xi^{0}_{\Gamma \vk, \vq}$ and $\Xi^0_{\Gamma S, \vq}$
  to denote an individual $\vq$-point's contribution to the IEHPP self-energy.
The exciton-phonon coupling matrix elements are
\begin{align}
    g_{S S'}(\vQ, \vq)
    & = \sum_{\vk}   A^{S*}(\vk,\vQ+\vq) A^{S'}(\vk,\vQ)
                      g_{c}\nonumber\\
    & -            A^{S*} (\vk,\vQ+\vq) A^{S'}(\vk+\vq, \vQ)
                      g_{v}
\end{align}
and the exciton self-energy is given by
\begin{align}
  \Xi_{\Gamma S} (\omega, T)
  = & \sum_{\vq} \sum_{S'}^{N} \frac{\vert g_{SS'}(\Gamma,\vq) \vert^2 P_{\pm}(T)}
                      {\omega - \Omega_{\vq S'} \pm \omega_0 + i\eta}
.
\end{align}

In general, the completion term for the self-energy
  contains the contribution of the valence and conduction bands excluded
  from the two-particle basis,
  as well as the contribution of the exciton states
  that are not explicitly computed.
Within this model, all the electron bands are
  included in the basis, but only the first $N$
  exciton bands are computed.
The completion term thus corresponds
  to the contribution of the remaining
  exciton states ($S'>N$).
We express the completion term in the form~\cite{SM}
\begin{align}
  \Xi_{\Gamma S}^{\text{C}} (\omega, T)
  = & \sum_{\vq} \bigg(
    1 -
    \frac{\tilde{\zeta}_{\Gamma S}(\vq)}{\zeta_{\Gamma S}(\vq)}
    \bigg)
    \Xi^{0}_{\Gamma S , \vq} (\omega,T)
,
\end{align}
where we define the partial sum
\begin{align}
    \tilde{\zeta}_{\Gamma S}(\vq) = 
    \sum_{S'=1}^{N} \vert g_{SS'}(\Gamma,\vq) \vert^2
\end{align}
and the corresponding summation over all exciton
  states can be obtained with the sum rule
\begin{align}
  \zeta_{\Gamma S}(\vq)
  =& \sum_{S'} \vert g_{SS'}(\Gamma, \vq) \vert^2
\\
  =& \vert g_c \vert^2 + \vert g_{v} \vert^2
   - 2 g_v^* g_c \sum_{\vk} A^{S *}(\vk + \vq, \Gamma) A^S(\vk, \Gamma)
\nonumber
.
\end{align}
The ratio
  $\tilde{\zeta}_{\Gamma S}(\vq) / \zeta_{\Gamma S}(\vq)$
  thus describes how much the first $N$ excitons
  probe the space of available states
  at wavevector $\vq$
  for the exciton $\Gamma S$ to couple with.

\subsection{Results and discussion}

Figure \ref{fig:OmegaT}(a-b) presents the $\vq$-space convergence
  of the imaginary part of the exciton-phonon self-energy.
The use of a finite value for the infinitesimal parameter $\eta$
  eases the convergence with respect to the number of $\vq$-points,
  but also introduces
  an arbitrarily small error in the lifetimes.
For the first optical exciton ($S=1$ in the exciton labeling),
  the inverse lifetime due to phonons is known to be zero at $T\!=\!0$, 
  since this exciton cannot scatter into a lower energy
  states by phonon emission.
The finite value obtained for $\Im \Xi_{S}$  %$\tau_1^{-1}$
  with a converged $\vq$-grid
  thus indicates the magnitude of the error.
A value of $\eta=10$~meV
  yields an error smaller than $3$~meV
  for the self-energy,
  and we use this value for the following 
  computations of $\Xi$.
For the second optical exciton state ($S=2$),
  we conclude %from Figure~\ref{fig:OmegaT}
  that a $48\times 48$ $\vq$-grid is well converged.
The temperature-dependent energy shift
  and the spectral width
  for the seven lowest
  exciton states are presented in Fig.~\ref{fig:OmegaT}c.
The lowest exciton state, having the largest binding energy,
  consequently has a longer lifetime than the others at all temperatures.

\begin{figure}[t]
  \includegraphics[width=1.0\linewidth]{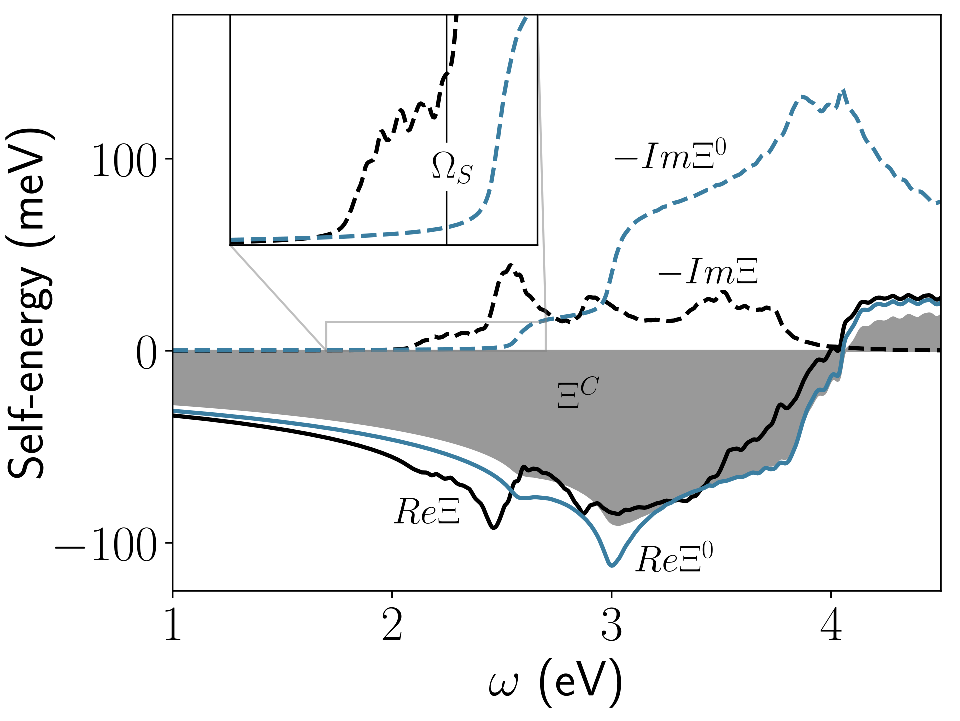}
  \caption{\label{fig:XiOmega}
  Real (solid lines) and imaginary part (dashed lines) of the
  exciton-phonon self-energy
  with (black) and without (blue) electron-hole interaction.
  The height of the filled
  region is the contribution of the completion term
  to the interacting self-energy.
}
\end{figure}

Figure \ref{fig:XiOmega} presents the frequency-dependent self-energy
  for the second lowest bound exciton.
The height of the 
  shaded area represents the completion term,
  which accounts for a large fraction of the real part of $\Xi$
  at all frequencies.
Near the
  $S\!=\!2$
  exciton energy ($\Omega_S\!=\!2.4$~eV),
  the imaginary part of $\Xi^0$ doesn't
  have any structure, since, without electron-hole interactions,
  the electron-hole pairs lie above the band gap.
However, the presence of bound excitons
  with finite crystal momentum near $\Gamma$ and $K$
  allow the optical excitons to scatter and diffuse,
  confering a finite value to the imaginary part of $\Xi$.
Just as the electron-hole interaction
  binds the excitons below the band gap,
  it also moves the spectral weight of the self-energy
  towards lower frequencies.

\begin{figure}[t]
  \includegraphics[width=\linewidth]{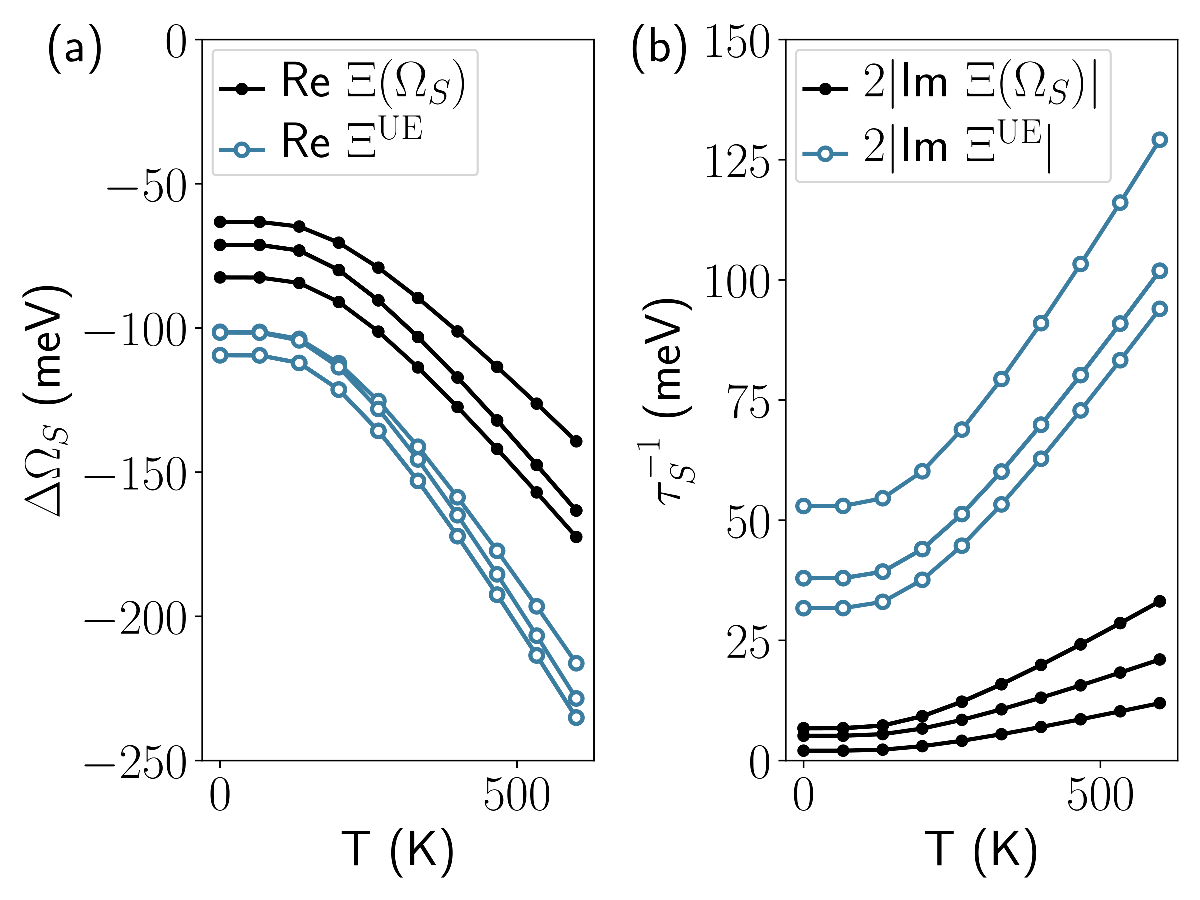}
  \caption{\label{fig:XiComparison}
  Comparison between the exciton-phonon self-energy (black discs)
  and an approximate expression (blue circles) for the first four bright excitons.
  \textbf{a\textendash} Energy shift.
  \textbf{b\textendash} Inverse lifetime.
}
\end{figure}

Let us now evaluate the accuracy
  of previously-used approximate expressions
  for the real and imaginary parts of the self-energy.
The
  uncorrelated exciton
  approximation of Eq.~\eqref{XiFXA} corresponds to writing 
\begin{align}
    \Xi^{\text{UE}}_{\Gamma S}
    =  \sum_{\vk} \vert A^S(\vk,\Gamma) \vert^2 \Xi^0_{\Gamma\vk}(\eig_{\vk c} - \eig_{\vk v})
.
\end{align}
Unlike the full exciton-phonon self-energy,
  this expression only requires the computation of the exciton's
  wave function for the state $\Gamma S$.
Figure~\ref{fig:XiComparison},
  compares the real and imaginary parts of $\Xi^{\text{UE}}$
  to those of $\Xi$.
The uncorrelated exciton approximation
  overestimates the inverse lifetime (broadening)
  by an order or magnitude.
For the phonon-induced energy shift $\Delta \Omega_S$,
  the overestimation of the
  negative shift (renormalization)
  for the seven lowest optical excitons
  is about 20\% to 40\%.

\begin{figure}[t]
  \includegraphics[width=\linewidth]{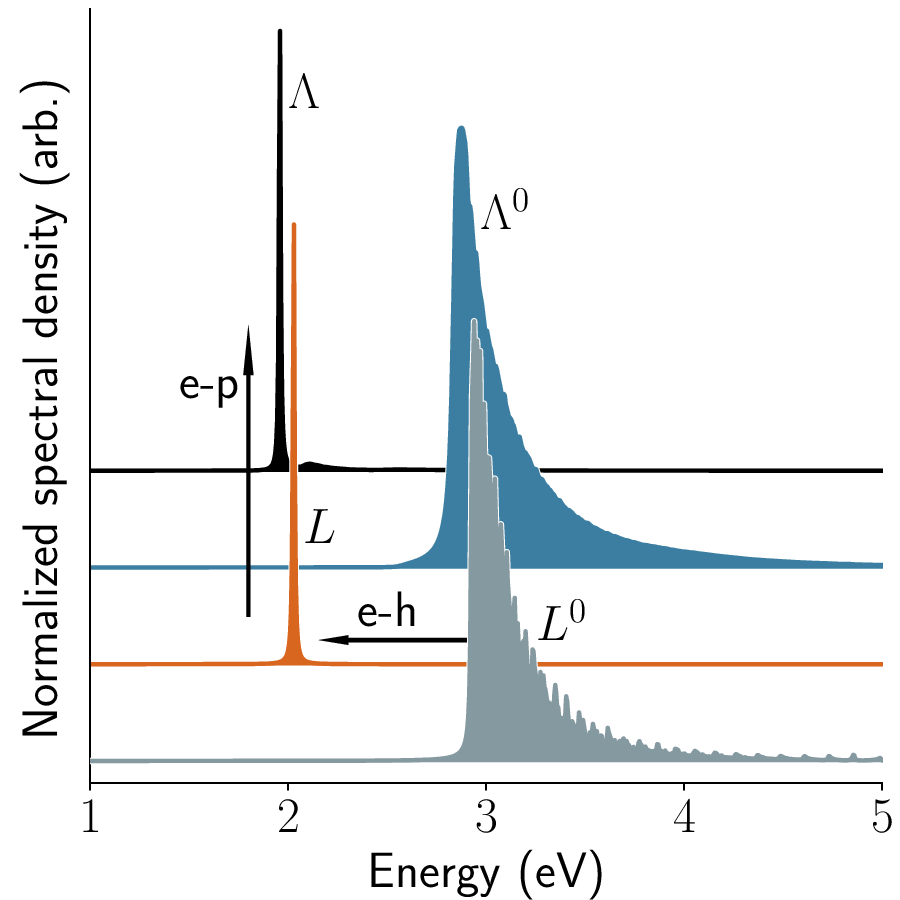}\\
  \caption{\label{fig:SF}
  Spectral function (imaginary part of the propagator)
  for the first bright exciton at $T\!=\!0$
  computed with the model hamiltonian
  at different levels of theory
  (in all cases, the imaginary part of the self-energy of the particle
  due to electron-electron interaction is neglected):
  non-interacting (gray),
  electron-hole interaction only (orange),
  electron-phonon interaction only (blue),
  and both interactions (black).
  }
\end{figure}

The combined effect of electron-hole and electron-phonon interactions
  is summarized in Fig.~\ref{fig:SF},
  which shows the exciton propagator for the lowest bound exciton
  as the electron-hole and electron-phonon interactions
  are switched on separately ($L$, $\Lambda^0$) or simultaneously ($\Lambda$).
In all cases, the imaginary part of the self-energy of the particles
  due to electron-electron interaction is neglected.  
An artificial broadening parameter ($\eta\!=\!5$~meV)
  is used to represent the bare exciton propagator ($L$),
  which would otherwise be infinitely sharp
  around the exciton energy $\Omega_S$
  because of the neglect of the lifetime of the particles
  due to electron-electron interaction
.
The function $L^0(\omega)$ illustrates the spectral decomposition of $L$
  into electron-hole pairs.
The spiky features in $L^0$ are an artifact
  of the finite $\vk$-points sampling used,
  and the function can be made smooth
  by using a larger broadening parameter $\eta$.
Turning on the electron-phonon interaction
  broadens these features,
  as can be seen by comparing $L^0$ with $\Lambda^0$.
The exciton propagator in the presence of the phonons field ($\Lambda$)
  is red-shifted with respect to the bare exciton propagator ($L$),
  and a satellite peak appears
  above the
  exciton peak in $\Lambda$.

\section{Conclusion}
\label{sec:Conclusion}

In summary, we have derived
  a rigorous
  expression for the exciton-phonon
  coupling self-energy to lowest order in the electron-phonon interaction
  and in the limit of low exciton density.
Through the exciton-phonon coupling matrix elements,
  the optically accessible excitons 
  may scatter into optically dark finite-momentum exciton states,
  resulting in
  an energy renormalization and 
  a finite lifetime for the optical excitations.
Our expressions takes into account the electron-hole interaction,
  and improve upon approximate expressions found in the literature
  by naturally enforcing energy conservation.

We implemented this theory on a two-dimensional two-band model
  and computed the temperature-dependent energy shift
  and lifetimes of the optical excitons.
This model allowed us to compare our exciton-phonon self-energy
  with an approximate expression which
  we call the uncorrelated exciton (UE) approximation.
We showed that the previously used approximation
  overestimates the inverse lifetime by an order of magnitude,
  making this approximation unreliable.
We conclude that, in physical systems
  with strong electron-hole interaction
  such as low-dimensional materials,
  it is necessary to use the exact
  exciton-phonon coupling theory
  to compute accurately the lifetime of optical excitations.
The UE approximation also overestimates
  the temperature-dependent shift of the energy of the excitons
  by 20-40\% in our two-band model.
We interpret this result as an upper bound
  for the error induced by this approximation
  in realistic systems.

The scheme developed in this paper readily applies
  to the study of exciton diffusion dynamics.
A main challenge in applying this theory is
  the computation of finite-momentum excitons
  energies and wavefunctions.
While such calculation has been demonstrated,
  a full sampling of the Brillouin zone
  remains computationally expensive
  and would benefit from interpolation techniques~%
  \cite{%
  qiu_nonanalyticity_2015,%
  berghauser_analytical_2014,%
  selig_dark_2018%
  }%
  .

\begin{acknowledgments}

G.A. acknowledges
  valuable
  discussions with
  Jonah B. Haber,
  Sivan Refaely-Abramson,
  Diana Y. Qiu,
  and
  Felipe H. da~Jornada.

\smallskip

This research was supported by
  the National Science Foundation under Grant No. DMR-1926004
  (conceptual and formalism development),
  by the Center for Computational Study of Excited-State Phenomena
  in Energy Materials (C2SEPEM) funded by the U. S. Department of Energy,
  Office of Basic Energy Sciences, under Contract No. DE-AC02-05CH11231
  at Lawrence Berkeley National Laboratory
  (code development),
  and by the Natural Sciences and Engineering Research Council of Canada
  under Grant No. RGPIN-2019-07149
  (computation).
\end{acknowledgments}

\end{document}

% Document info
\title{Theory of the Exciton-Phonon Coupling: Supplemental information}
\author{Gabriel Antonius} \email{\uqtrmail}
\affiliation{\uqtr}
\author{Steven G. Louie}
\affiliation{\berkeley}

\affiliation{\berkeley}

\maketitle

\section{Quasiparticle propagators and frequency summations} \label{sec:Matsubara}

This section presents the Matsubara formalism used
  and a few important Matsubara summations
  that lead to the analytic expressions
  of the different self-energies
  presented in the manuscript.
The transform between imaginary time and imaginary frequency
  for a quantity $A$ (either a propagator or a self-energy) is
\begin{align}
  A(\tau) = & \frac{1}{\beta} \sum_{n=-\infty}^{\infty}
    e^{-i\omega_n \tau} A(i\omega_n)
  \qquad ; \qquad
  A(i\omega_n) =  \int_0^{\beta} d\tau
    e^{i\omega_n \tau} A(\tau)
\end{align}
where $\beta = 1/k_B T$,
and the Matsubara frequencies are defined as
\begin{align}
  \omega_n = & \frac{2n\pi}{\beta} \quad \text{(bosons)}
  \qquad ; \qquad %\\
  \omega_m =  \frac{(2m+1)\pi}{\beta} \quad \text{(fermions)}
\end{align}
Throughout the paper,
  we use $i\omega_n$ and $i\omega_l$ for bosonic frequencies,
  and $i\omega_m$ for fermionic frequencies.
We will recover all the retarded quantities on the real frequency axis
  using a real positive infinitesimal number $\eta$
  and the analytic continuation
\begin{align}
  A^R(\omega) = \lim_{i\omega_n \rightarrow \omega + i\eta} A(i\omega)
.
\end{align}
In this formalism,
  the one-particle propagator is
\begin{equation}
  G^0_{ii'}(i\omega_m) = \frac{1}{i\omega_m - \eig_i} \delta_{ii'}
\end{equation}
while the phonon propagator is
\begin{equation}
  D_\ipho(i \omega_l) = \frac{1}{i\omega_l - \omega_\ipho}
                  - \frac{1}{i\omega_l + \omega_\ipho}
\end{equation}
Note that $D_\ipho(i\omega_l)=D_\ipho(-i\omega_l)$.
The bare electron-hole propagator is
\begin{align}
  L^0_{vcv'c'}(i\omega_n) = \frac{f(\eig_v) - f(\eig_c)}
                                 {i\omega_n - (\eig_c - \eig_v)}
                            \delta_{cc'} \delta_{vv'}
\end{align}
And the bare exciton propagator is
\begin{align}
 L_{SS'}(i\omega_n) = \frac{1}{i\omega_n - \Omega_S}\delta_{SS'}
\end{align}
For the phonon propagator,
  the summation over bosonic Matsubara frequencies yields
\begin{align} \label{sumD}
  - \frac{1}{\beta} \sum_{l} D_{\ipho}(i\omega_l)
    = 2 N_B(\omega_\ipho) + 1
\end{align}
where $N_B(\omega)$ is the Bose-Einstein distribution defined as
\begin{align}
  N_B(\omega) = \frac{1}{e^{\beta \omega} - 1}
\end{align}
For the one-particle propagator,
  the summation over fermionic Matsubara frequencies yields
\begin{align} \label{sumG}
  \frac{1}{\beta} \sum_{m} G^0_{ii}(i\omega_m) = f(\eig_i)
\end{align}
where $f(\omega)$ is the Fermi-Dirac distribution defined as
\begin{align}
  f(\omega) = \frac{1}{e^{\beta \omega} + 1}
\end{align}
assuming that $\omega$ is measured from the chemical potential $\mu$.
The bare electron-hole propagator is expressed
  as the convolution of two one-particle propagators
\begin{align} \label{LGG}
  L^0_{vc,vc}(i\omega_n)
  = \frac{1}{\beta} \sum_{m} G^0_{vv}(i\omega_m) G^0_{cc}(i\omega_m + i\omega_n)
\end{align}
Upon computing the one-particle electron-phonon self-energy,
  we encouter a summation of the type
\begin{align} \label{sumDG}
  - \frac{1}{\beta} \sum_l &
   D_\ipho(i\omega_l) G^0_{ii}(i\omega_l + i\omega_m)
    = \frac{N_B(\omega_\ipho) + 1 - f(\eig_i)}
           {i\omega_m - \eig_i - \omega_\ipho}
    + \frac{N_B(\omega_\ipho) + f(\eig_i)}
           {i\omega_m - \eig_i + \omega_\ipho}
\end{align}
while,
  for the IEHPP self-energy $\Xi^0$,
  we encouter summations of the type
\begin{align} \label{sumDL}
  - \frac{1}{\beta} \sum_l &
   D_\ipho(i\omega_l) L^0_{vc,vc}(i\omega_l + i\omega_n)
\nonumber\\
  & = \big(f(\eig_v) - f(\eig_c)\big) \bigg[
        \frac{N_B(\omega_\ipho) + 1 + n(\eig_c - \eig_v)}
             {i\omega_n - (\eig_c - \eig_v) - \omega_\ipho}
      + \frac{N_B(\omega_\ipho) - n(\eig_c - \eig_v)}
             {i\omega_n - (\eig_c - \eig_v) + \omega_\ipho}
          \bigg]
\nonumber\\
  & \approx 
      \frac{N_B(\omega_\ipho) + 1}
           {i\omega_n - (\eig_c - \eig_v) - \omega_\ipho}
    + \frac{N_B(\omega_\ipho)}
           {i\omega_n - (\eig_c - \eig_v) + \omega_\ipho}
\end{align}
The last approximation results from the large-band-gap approximation,
  which is explained in appendix \ref{sec:LBGA}.

\clearpage
\section{Large band gap approximation} \label{sec:LBGA}

In order to arrive at an expression for all the contributions
  to $\Xi^0$, we will take advantage of the large-band-gap approximation
  for the one-particle occupation numbers, namely
\begin{align}
  f(\eig_v) \approx 1 \quad \forall \ v\\
  f(\eig_c) \approx 0 \quad \forall \ c
\end{align}
For materials with band gaps larger than 1 eV,
  this approximation holds for temperatures up to $10, 000$~K.
%This condition holds for bandgap
As a result, the bare electron-hole propagator
  writes
\begin{align}
  L^0_{vcv'c'}(i\omega_n) \approx \frac{1}{i\omega_n - (\eig_c - \eig_v)}
                            \delta_{cc'} \delta_{vv'}
\end{align}
And we have the identity
\begin{align}
  G^0_{ii}(i\omega_m)G^0_{jj}(i\omega_m + i\omega_n) =
    L^0_{ij,ij}(i\omega_n)
    \big[
      G^0_{ii}(i\omega_m)
      - G^0_{jj}(i\omega_m + i\omega_n)
    \big]
\end{align}
Using this result, we can show that
\begin{align} \label{threeGexpanded}
 G^0_{vv}(i\omega_m) & G^0_{v'v'}(i\omega_m) G^0_{cc'}(i\omega_m + i\omega_n)
    = L^0_{vc,vc}(i\omega_n) L^0_{v'c',v'c'}(i\omega_n) \delta_{cc'}\nonumber\\
    \times \bigg\{ &  G^0_{cc'}(i\omega_m + i\omega_n)
                    - \frac{1}{2} \big[ G^0_{vv}(i\omega_m)
                                      + G^0_{v'v'}(i\omega_m) \big]\nonumber\\
      & + \frac{1}{2} \big[ L^{0^{-1}}_{vc,vc}(i\omega_n)
                          + L^{0^{-1}}_{v'c',v'c'}(i\omega_n) \big]
 G^0_{vv}(i\omega_m) G^0_{v'v'}(i\omega_m)
    \bigg\}
\end{align}
which has been symmetrized in $(v,v')$.
The third line of Eq.~\eqref{threeGexpanded} gives a contribution
  that is exactly zero if it is summed along a frequency-independent term,
  such as the DW self-energy.
Its contribution is also zero when evaluated at $i\omega_n = \eig_c - \eig_v$,
  and is vanishing in the neighborhood of this value.
It will thus be neglected.
To simplify the notation, we introduce the symmetrization symbol $\sym$,
  which, for a quantity $A_{vc,v'c'}$, means
\begin{align}
  \sym \left[ A_{vc,v'c'} \right]
  = \frac{1}{2} \big( A_{vc,v'c'} + A_{v'c',vc} \big)
.
\end{align}
Quantities with two or three indices are obtaind by a contraction
  of indices and symmetrize as
%and for a quantity with three indices,
\begin{align}
  \sym \left[ A_{vcc'} \right] = \frac{1}{2} \big( A_{vcc'} + A_{vc'c} \big)
  \quad ; \quad
  \sym \left[ A_{vv'c} \right] = \frac{1}{2} \big( A_{vv'c} + A_{v'vc} \big)
  \quad ; \quad
  \sym \left[ A_{ii'} \right] = \frac{1}{2} \big( A_{ii'} + A_{i'i} \big)
\end{align}
Hence, we arrived at the approximation
\begin{align} \label{LBAtwo}
  G^0_{vv}(i\omega_m)  G^0_{v'v'}(i\omega_m) G^0_{cc'}(i\omega_m + i\omega_n)
   & \approx L^0_{vc,vc}(i\omega_n) L^0_{v'c',v'c'}(i\omega_n)\nonumber\\
   & \times \sym \big[   G^0_{cc'}(i\omega_m + i\omega_n)
                    - G^0_{vv}(i\omega_m) \big] \delta_{cc'}
\end{align}
A similar reasonning gives
\begin{align} \label{LBAthree}
  G^0_{cc}(i\omega_m + i\omega_n) 
  G^0_{c'c'}(i\omega_m + i\omega_n) G^0_{vv'}(i\omega_m)
  & \approx L^0_{vc,vc}(i\omega_n) L^0_{v'c',v'c'}(i\omega_n)\nonumber\\
  & \times \sym \big[  G^0_{vv'}(i\omega_m)
                    - G^0_{cc}(i\omega_m + i\omega_n) \big] \delta_{vv'}
\end{align}
Another result from the large-band-gap approximation is
\begin{align} \label{LBAfour}
  & G^0_{cc}(i\omega_n + i \omega_m) G^0_{vv}(i\omega_m)
    G^0_{c'c'}(i\omega_n + i\omega_m - i\omega_l)
    G^0_{v'v'}(i\omega_m - i \omega_l)\nonumber\\
  & = L^0_{vc,vc}(i\omega_n) L^0_{v'c',v'c'}(i\omega_n)
           \big[ G^0_{vv}(i\omega_m)
               - G^0_{cc}(i\omega_n + i \omega_m) \big]\nonumber\\
  & \times \big[ G^0_{v'v'}(i\omega_m - i \omega_l)
               - G^0_{c'c'}(i\omega_n + i\omega_m - i\omega_l) \big]
\end{align}
and
\begin{align} \label{LBAfive}
  \sum_m & \big[ G^0_{vv}(i\omega_m)
             - G^0_{cc}(i\omega_n + i \omega_m) \big]
        \big[ G^0_{v'v'}(i\omega_m - i \omega_l)
             - G^0_{c'c'}(i\omega_n + i\omega_m - i\omega_l) \big]\nonumber\\
 &\approx - \big[ L^0_{v'c,v'c}(i\omega_n + i\omega_l)
                + L^0_{vc',vc'}(i\omega_n - i\omega_l) \big]
\end{align}
Finally, we assume that all phonon frequencies are much smaller than
  the band gap and the exciton energies.
This approximation allow us to write
\begin{align}
  f(\eig_i \pm \omega_{\ipho}) \approx f(\eig_i)
\end{align}
and
\begin{align}
  N_B(\omega_{\ipho}) \gg & N_B(\eig_c - \eig_v)\nonumber\\
  N_B(\omega_{\ipho}) \gg & N_B(\Omega_S)
\end{align}
for any pair of valence and conduction states
  with energies $\eig_v$ and $\eig_c$),
  any exciton energy $\Omega_S$,
  and any phonon energy $\omega_{\ipho}$.

\clearpage
\section{More detailed derivation of $\Xi^0$} \label{sec:MoreDetailedXi0}

We proceed to a detailed derivation of the IEHPP
  self-energy $\Xi^0$ by expanding $\Lambda^0$
  from Eq.~\eqref{M-Lambdazexpandone}.
To the lowest order in the perturbation,
  the self-energy will be given by
\begin{equation} \label{XiZeroFirstOrder}
  \Xi^0_{vcv'c'}(i\omega_n) =
    L^{0^{-1}}_{vc,vc}(i\omega_n)
    \Big[ \Lambda^0_{vc,v'c'}(i\omega_n) - L^0_{vc,v'c'}(i\omega_n) \Big]
    L^{0^{-1}}_{v'c',v'c'}(i\omega_n)
\end{equation}
Hence, each term in $\Lambda^0$ beyond $L^0$ contributes to $\Xi^0$.

The first set of terms are called the Debye-Waller terms,
  since they involve the second-order electron-phonon coupling potential.
They write
\begin{align}
  \Lambda^{0 DW}_{vc,v'c'}(i\omega_n) =
&  \Sigma^{DW}_{c'c} \Big[ \frac{1}{\beta} \sum_m
     G^0_{cc}(i\omega_m) G^0_{c'c'}(i\omega_m)
     G^0_{vv'}(i\omega_m - i\omega_n)
   \Big]\nonumber\\
+ & \Sigma^{DW}_{vv'} \Big[ \frac{1}{\beta} \sum_m
     G^0_{vv}(i\omega_m) G^0_{v'v'}(i\omega_m)
     G^0_{cc'}(i\omega_m + i\omega_n)
   \Big]
\end{align}
where
\begin{align} \label{DWtwoep}
  \Sigma^{DW}_{ii'} = - \sum_{\ipho} g^{(2)}_{ii' \ipho \ipho }
                        \frac{1}{\beta} \sum_{l} D_{\ipho}(i\omega_l)
\end{align}
We recover Eq.~\eqref{M-DWonep} for $\Sigma^{DW}$
   by making use of Eq.~\eqref{sumD}.
The next set of terms are called the Fan-Migdal (FM) terms.
They write
\begin{align} \label{Lambda_Fan}
  \Lambda^{0 FM}_{vc,v'c'}(i\omega_n) = - \frac{1}{\beta} \sum_m \bigg\{
&  \Sigma^{FM}_{cc'}(i\omega_m) \Big[
     G^0_{cc}(i\omega_m) G^0_{c'c'}(i\omega_m)
     G^0_{vv'}(i\omega_m - i\omega_n)
   \Big]\nonumber\\
+ & \Sigma^{FM}_{v'v}(i\omega_m) \Big[
     G^0_{vv}(i\omega_m) G^0_{v'v'}(i\omega_m)
     G^0_{cc'}(i\omega_m + i\omega_n)
   \Big]\bigg\}
\end{align}
where
\begin{align} \label{Fantwoep}
  \Sigma^{FM}_{ii'}(i\omega_n) = &
    - \sum_{i'' \ipho}
    g_{ii'' \ipho} g^{*}_{i' i'' \ipho}
    \frac{1}{\beta}\sum_l
    D_\ipho(i\omega_l) G^0_{i''}(i\omega_n + i\omega_l)
\end{align}
We recover Eq.~\eqref{M-Fanonep} for $\Sigma^{FM}$
  by making use of Eq.~\eqref{sumDG}.
The last set of terms are called the phonon exchange terms.
They write
\begin{align} \label{Lambda_X}
  \Lambda^{0^X}_{vc,v'c'}(i\omega_n) = - \sum_{\ipho} g^{}_{c'c \ipho} g^{*}_{v'v\ipho}
    \frac{1}{\beta^2} \sum_{lm}
   &  G^0_{cc}(i\omega_n + i \omega_m)
      G^0_{c'c'}(i\omega_n + i\omega_m - i\omega_l)\nonumber\\
\times &
      G^0_{v'v'}(i\omega_m - i \omega_l) G^0_{vv}(i\omega_m)
      D_{\ipho}(i\omega_l)
\end{align}

Using Eq.~\eqref{threeGexpanded} and Eq.~\eqref{LBAtwo},
  the Debye-Waller contribution to $\Xi^0$ is
\begin{align}
   \Xi^{0^{DW}}_{vc,v'c'} =
   \frac{1}{\beta} \sum_m
  & \Sigma^{DW}_{v'v} \sym \Big[
    G^0_{cc'}(i\omega_m + i\omega_n)
  - G^0_{vv}(i\omega_m)
    \Big]\delta_{cc'}\nonumber\\
+ & \Sigma^{DW}_{cc'} \sym \Big[
    G^0_{vv'}(i\omega_m - i\omega_n)
  - G^0_{cc}(i\omega_m) \Big]\delta_{vv'}
\end{align}
Within the large-bandgap-approximation,
the Debye-Waller term is simply
\begin{align}
  \Xi^{0^{DW}}_{vc,v'c'} =
    \Sigma^{DW}_{cc'}\delta_{vv'}
  - \Sigma^{DW}_{v'v}\delta_{cc'} 
\end{align}

To obtain the FM term contribution,
  we insert Equations \eqref{Lambda_Fan} into \eqref{XiZeroFirstOrder},
  and use \eqref{LBAtwo} and \eqref{LBAthree},
  giving
\begin{align}
   \Xi^{0^{FM}}_{vc,v'c'}(i\omega_n) = 
    \frac{1}{\beta} \sum_m
  & \Sigma^{FM}_{v'v}(i\omega_m) \sym \Big[
    G^0_{cc'}(i\omega_m + i\omega_n)
  - G^0_{vv}(i\omega_m)
    \Big]\nonumber\\
+ & \Sigma^{FM}_{cc'}(i\omega_m) \sym \Big[
    G^0_{vv'}(i\omega_m - i\omega_n)
  - G^0_{cc}(i\omega_m) \Big]
\end{align}
We split the FM term into two contributions:
  the dynamical FM term (FMd) and the static FM term (FMs).
We define the dynamical Fan term as
\begin{align}
  \Xi^{0^{FMd}}_{vc,v'c'}(i\omega_n) =
    \frac{1}{\beta} \sum_m 
    \Sigma^{FM}_{v'v}(i\omega_m) G^0_{cc'}(i\omega_m + i\omega_n)
+   \Sigma^{FM}_{cc'}(i\omega_m) G^0_{vv'}(i\omega_m - i\omega_n)
\end{align}
and the static FM term as
\begin{align}
  \Xi^{0^{FMs}}_{vc,v'c'} = 
  - \frac{1}{\beta} \sum_m
    \Sigma^{FM}_{vv'}(i\omega_m) \sym G^0_{vv}(i\omega_m) \delta_{cc'}
+   \Sigma^{FM}_{cc'}(i\omega_m) \sym G^0_{cc}(i\omega_m) \delta_{vv'}
\end{align}
We use Eq.~\eqref{M-Fanonep}
  to express the one-particle self-energy
  as a convolution of the Green's functions of an electron and a phonon,
  and use Eq.~\eqref{LGG} to write
\begin{align}
  \Xi^{0^{FMd}}_{vcv'c'}& (i\omega_n) =
  - \frac{1}{\beta} \sum_l \sum_{\ipho} 
    \sum_{v''c''} D_\ipho(i\omega_l)
      L^0_{v''c''v''c''}(i\omega_l + i\omega_n)\nonumber\\
  &
    \times \Big[  g_{v'v'' \ipho} g^{*}_{v v'' \ipho}
                   \delta_{cc''}\delta_{c''c'}
                + g_{cc'' \ipho} g^{*}_{c' c'' \ipho} \delta_{vv''}
                   \delta_{v''v'}
     \Big]
\end{align}
and
\begin{align}
  \Xi^{0^{FMs}}_{vcv'c'}
  & = \frac{1}{\beta} \sum_l \sum_{\ipho}
    \sum_{v''c''} D_\ipho(i\omega_l) L^0_{v''c''v''c''}(i\omega_l)\nonumber\\
   &
    \times \Big[ g_{v'c'' \ipho} g^{*}_{v c'' \ipho} \delta_{cc'}  
                 \sym \delta_{vv''}
         + g_{cv'' \ipho} g^{*}_{c' v'' \ipho} \delta_{vv'}
            \sym \delta_{cc''}
     \Big]
\end{align}

To obtain the phonon exchange term contribution,
  we insert Equations \eqref{Lambda_X} into \eqref{XiZeroFirstOrder},
  and use \eqref{LBAfour} and \eqref{LBAfive},
  giving
\begin{align}
  \Xi^{0^{X}}_{vcv'c'}(i\omega_n) = g_{c'c \ipho} g^{*}_{v' v \ipho}
    \frac{1}{\beta} \sum_l \big[ L^0_{vc',vc'}(i\omega_n + i\omega_l)
                               + L^0_{v'c,v'c}(i\omega_n + i\omega_l) \big]
                           D_{\ipho}(i\omega_l)
\end{align}
Performing the imaginary frequency summation,
  we obtain Eq.~\eqref{M-Xi0_X_final}. % FIXME

\section{Completion term within the two-band model}
\label{sec:modcomp}

\begin{figure}[h]
  \includegraphics[width=0.45\linewidth]{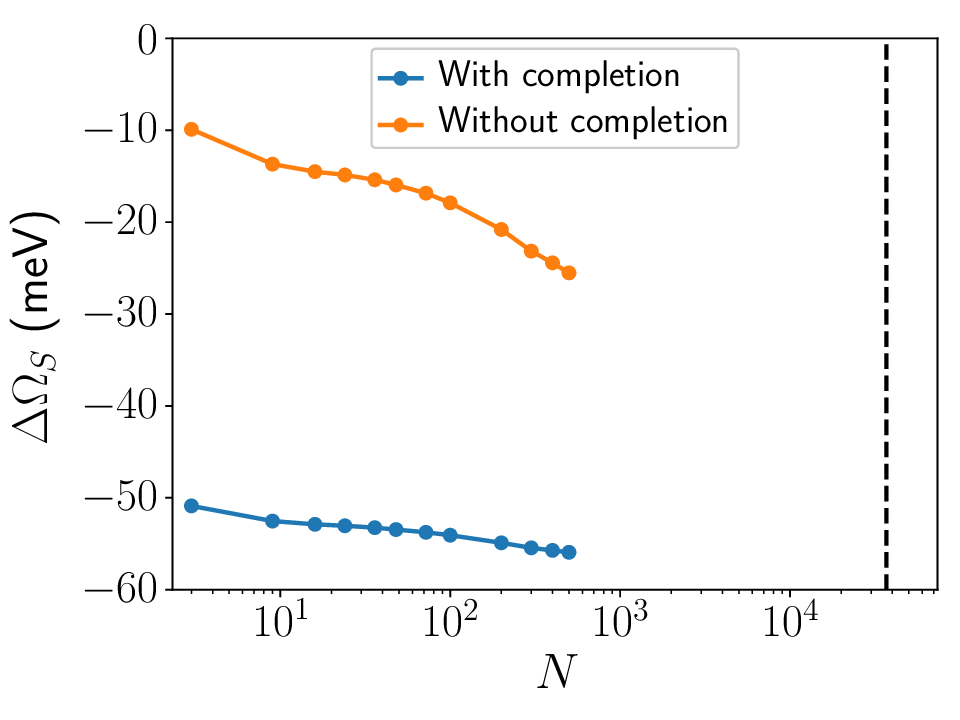}
  \caption{
  \label{fig:XiConvBands}
  Convergence of the exciton-phonon self-energy
  for $S=0$
  with respect to the number of exciton bands,
  with (blue) and without (orange) the completion term.
  The dashed line represents the limit of complete basis
  ($N=N_k$)
  where both lines are expected to meet.
}
\end{figure}

In the present calculation,
  the wavevector basis allows for a maximum of $N_k$ exciton states,
  but the first $N$ exciton states are actually computed,
The exciton-phonon self-energy therefore includes a completion term
    to account for the missing transitions.
Let $\ket{\vk,\vq}$ denote the state with one hole at wavevector $\vk$
  and one electron in the conduction band at wavevector $\vk+\vq$,
  such that $\braket{\vk,\vq \vert S,\vq} = A^{S}(\vk, \vq)$.
The electron-phonon coupling operator is
\begin{align}
  V_{ep}(\vq) = \sum_{\vk} g_c \ket{\vk, \vq} \bra{\vk, \Gamma} - g_v \ket{\vk-\vq, \vq} \bra{\vk,\Gamma}
\end{align}
We complete the self-energy using the equivalent completion phase space
  of the non-interacting self-energy, that is
\begin{align*}
  \Xi^{0}_{k} (\Gamma, \omega)
  %= & \sum_{\vk''\vq} \frac{ g_{kk'',vv,cc}(\vq)  g^*_{k'k'',vv,cc}(\vq)}
  %                    {\omega - (\varepsilon_{\vk''+\vq c} - \varepsilon_{\vk'' v} ) - \omega_\ipho + i\eta}
  = & \sum_{\vk'\vq} \frac{ \bra{\vk, \Gamma} V_{ep}(\vq) \ket{\vk', \vq}
                             \bra{\vk', \vq} V^*_{ep}(\vq) \ket{\vk, \Gamma}}
                      {\omega - (\varepsilon_{\vk'+\vq c} - \varepsilon_{\vk' v} ) \pm \omega_0 + i\eta}
                    P_{\pm}(T)
  = \sum_{\vq} \Xi^{0}_{\vk} (\Gamma, \omega, \vq)
\end{align*}
In the exciton basis, the non-interacting self-energy is non-diagonal,
  but we consider a diagonal element ($S,S$)
\begin{align*}
  \Xi^{0}_{SS} (\Gamma, \omega)
  = & \sum_{\vk',\vq} \frac{ g_{SS'}(\Gamma, \vq) g_{S S''}^*(\Gamma,\vq)
                A^{S'*}(\vk', \vq)
                A^{S''}(\vk', \vq)
                }
                {\omega - (\varepsilon_{\vk'+\vq c} - \varepsilon_{\vk' v} ) \pm \omega_0 + i\eta}
                P_{\pm}(T)
\end{align*}
Given that
\begin{align}
  \sum_{\vk} A^{S'*}(\vk', \vq) A^{S''}(\vk', \vq) = \delta_{S', S''}
\end{align}
in the limit of weak exciton binding,
  $\Xi^0_{SS}$ must correspond to $\Xi_{SS}$, namely
\begin{align*}
  \Xi_{SS} (\vQ, \omega, T)
  = & \sum_{\vq} \sum_{S'} \frac{\vert g_{SS'}(\vQ,\vq) \vert^2}
                      {\omega - \Omega_{S'} \pm \omega_0 + i\eta}
                      P_{\pm}(T)
\end{align*}
The completion term is the missing terms, when only the first $N$
  excitons are included in the basis, that is
\begin{align*}
  \Xi_{SS}^{C} (\vQ, \omega, T)
  = & \sum_{\vq}
  \bigg\{
      \sum_{S'} \frac{\vert g_{SS'}(\vQ,\vq) \vert^2 P_{\pm}(T)}
                      {\omega - \Omega_{S'} \pm \omega_0 + i\eta}
    - \sum_{S'=1}^{N} \frac{\vert g_{SS'}(\vQ,\vq) \vert^2 P_{\pm}(T)}
                      {\omega - \Omega_{S'} \pm \omega_0 + i\eta}
  \bigg\}
\end{align*}
We first approximate the contribution of the $S'$ states
  as those of electron-hole pairs
\begin{align*}
  \Xi_{SS}^{C} (\vQ, \omega, T)
  = & \sum_{\vq} \bigg\{
      \Xi^{0}_{SS} (\Gamma, \omega,\vq)
    - \sum_{S'=1}^{N} \frac{\vert g_{SS'}(\vQ,\vq) \vert^2 P_{\pm}(T)}
                      {\omega - \Omega_{S'} \pm \omega_0 + i\eta}
  \bigg\}
\end{align*}
Next, define a $\vq$-dependent effective energy parameter $E_{eff}(S,\vq)$
  that allows to write the completion term as
\begin{align} \label{xicompinter}
  \Xi_{SS}^{C} (\vQ, \omega, T)
  = & \sum_{\vq} \bigg\{
      \Xi^{0}_{SS} (\Gamma, \omega,\vq)
    - \sum_{S'=1}^{N} \frac{\vert g_{SS'}(\vQ,\vq) \vert^2}
                      {\omega - E_{eff}(S,\vq) + i\eta}
  \bigg\}
\end{align}
In the limit
  of complete basis ($N\!=\!N_k$),
  each pair of terms in Eq.~\eqref{xicompinter}
  must cancel perfectly.
In this limit, we have the sum rule
\begin{align}
  \zeta_{\Gamma S}(\vq) = \sum_{S'} \vert g_{S,S'}(\Gamma, \vq) \vert^2
  = \vert g_c \vert^2 + \vert g_{v} \vert^2
    - 2 g_v^* g_c A^{S *}(\vk + \vq, \Gamma) A^S(\vk, \Gamma)
\end{align}
Making use of this sum rule
\begin{align}
    \frac{1}{\omega - E_{eff}(S,\vq) + i\eta}
      =
      \frac{1}{\zeta_{\Gamma S}(\vq)} \Xi^{0}_{SS} (\Gamma, \omega,\vq)
\end{align}
We define the partial summation (when $N<N_k$)
\begin{align}
    \tilde{\zeta}_{\Gamma S}(\vq) = 
    \sum_{S'=1}^{N} \vert g_{SS'}(\Gamma,\vq) \vert^2
\end{align}
which allows us to write
\begin{align}
  \Xi_{SS}^{C} (\vQ, \omega, T)
  = & \sum_{\vq} \bigg(
    1 -
    \frac{\tilde{\zeta}_{\Gamma S}(\vq)}{\zeta_{\Gamma S}(\vq)}
    \bigg)
    \Xi^{0}_{SS} (\Gamma, \omega,\vq)
\end{align}
This is the final expression for the completion term
  of the exciton-phonon coupling self-energy.
As shown in Fig.~\ref{fig:XiConvBands},
  the completion term significantly speeds up the convergence
  of the self-energy
  with respect to the number of exciton bands.

%\bibliography{main.bib}